\documentclass[11pt]{article}
\usepackage{mathptmx}
\usepackage{amsmath}
\usepackage{amssymb,amsthm}
\usepackage{makecell}
\setcellgapes{6pt} 

\usepackage{authblk}
\usepackage[final]{graphicx}
\usepackage{color}
\usepackage[margin=0.75in]{geometry}
\usepackage{chngcntr}
\usepackage{indentfirst}
\usepackage{enumitem}
\usepackage{hyperref}
\usepackage[abspath]{currfile}
\usepackage{float}
\usepackage{bbm}
\usepackage{natbib}
\usepackage{filecontents}
\usepackage[toc]{appendix}
\usepackage{appendix}
\usepackage{apxproof}
\DeclareMathOperator*{\argmax}{arg\,max}
\DeclareMathOperator*{\argmin}{arg\,min}

\title{Distributionally Robust XVA via Wasserstein Distance: \\
 Wrong Way Counterparty Credit and Funding Risk}
\author{Derek Singh, \, Shuzhong Zhang}
\affil{Department of Industrial and Systems Engineering, University of Minnesota \\ singh644@umn.edu, \, zhangs@umn.edu}
\date{\vspace{-5ex}}

\theoremstyle{case}

\newtheorem{remark}{Remark}
\newtheorem*{theorem*}{Theorem}
\newtheoremrep{prop}{Proposition}[section]
\newtheoremrep{theorem}{Theorem}[section]
\newtheoremrep{lemma}[theorem]{Lemma}
\newtheoremrep{conj}[theorem]{Conjecture}
\newtheoremrep{corollary}{Corollary}[prop]

\providecommand{\keywords}[1]
{
  \small	
  \textbf{\textit{Keywords---}} #1
}

\usepackage[section]{placeins}

\begin{document}
\maketitle
\begin{abstract}
This paper investigates calculations of robust XVA, in particular, credit valuation adjustment (CVA) and funding valuation adjustment (FVA) for over-the-counter derivatives under distributional uncertainty using Wasserstein distance as the ambiguity measure. Wrong way counterparty credit risk and funding risk can be characterized (and indeed quantified) via the robust XVA formulations. The simpler dual formulations are derived using recent infinite dimensional Lagrangian duality results. Next, some computational experiments are conducted to measure the additional XVA charges due to distributional uncertainty under a variety of portfolio and market configurations. Finally some suggestions for future work are discussed.
\end{abstract}

\keywords{CCR, CVA, FVA, derivatives, distributional robust optimization, Wasserstein distance, Lagrangian duality}

\section{Introduction and Overview}
\subsection{Financial Markets Context and Background}
\indent An X-Value adjustment (XVA) is a generic term used to refer to various valuation adjustments, typically applied to over-the-counter (OTC) derivatives held by financial institutions. The first of the XVAs, and still one of the most significant, in terms of market exposure, is CVA. One of the more recent, and perhaps equally significant, exposures is FVA. Both of these XVAs have similar structure (unilateral, bilateral) and mathematical form for computation. Other XVAs include capital valuation adjustment (KVA) and margin valuation adjustment (MVA). Wrong way risk refers to adversely correlated moves in the market exposures and the counterparty spreads (e.g. credit, funding). It can materially affect the magnitude of the XVA adjustment. \par
\indent Credit valuation adjustment (CVA) represents the impact on portfolio market value due to counterparty default. Unilateral CVA can be represented mathematically as an integral of discounted expected positive exposure times (incremental) counterparty default probability. The market valuation is a function of counterparty credit risk, the underlying (market) risk factors that drive the portfolio valuation (and hence positive exposure), as well as the correlations between these market risk factors and the counterparty credit risk curves for a given portfolio. CVA is typically measured and reported at the counterparty level. \par
The ``other side" of unilateral CVA is unilateral debit valuation adjustment (DVA). This is the benefit to the firm, of its reduced liability, as measured by discounted expected negative exposure times firm default probability. As above, the market valuation is a function of firm credit risk, underlying market risk factors that drive portfolio valuation, and the correlations. Unilateral DVA can be represented mathematically as an integral of discounted negative exposure times (incremental) firm default probability. DVA is typically measured at the firm level. \par
Bilateral CVA represents the dual impact on portfolio market value due to counterparty default and firm default. Bilateral CVA can be represented mathematically as the difference between two integrals: (i) discounted expected positive exposure times (incremental) counterparty default probability prior to firm default, (ii) discounted expected negative exposure times (incremental) firm default probability prior to counterparty default. Bilateral CVA is typically measured and reported at the counterparty level, for a given firm. \par
\indent Funding valuation adjustment (FVA) represents the impact on portfolio market value due to funding exposures for the hedge on uncollateralized derivatives. It represents the market value of funding exposure risk. Funding cost adjustment (FCA) can be represented mathematically as an integral of discounted expected positive exposure times funding cost (incremental) conditional on joint counterparty and firm survival. FCA arises for a positive portfolio exposure since this implies a negative hedge exposure which leads to a funding cost for collateral posted. The market valuation is a function of joint counterparty and firm credit risk, the underlying (market) risk factors that drive the portfolio valuation (and hence positive exposure) as well as funding cost, as well as the correlations between these market risk factors and the credit risk curves for a given portfolio. FCA is typically measured and reported at the funding netting set level. \par
The ``other side" of FCA is funding benefit adjustment (FBA). This represents the funding benefit to the firm, for interest income proceeds on received collateral posted against counterparty exposure on the hedge, as measured by discounted expected negative exposure times funding benefit conditional on joint counterparty and firm survival. As above, the market valuation is a function of counterparty and firm credit risk, underlying market risk factors that drive portfolio valuation and funding benefit, and the correlations. FBA can be represented mathematically as an integral of discounted negative exposure times funding benefit  conditional on joint counterparty and firm survival. FBA is typically measured at the funding netting set level. \par
(Bilateral) FVA represents the dual impact on portfolio market value due to both funding cost and funding benefit exhibited over the portfolio lifetime. FVA can be represented mathematically as the difference (or sum) of two integrals: (i) discounted expected positive exposure times funding cost conditional on joint counterparty and firm survival; (ii) discounted expected negative exposure times funding benefit conditional on joint counterparty and firm survival. FVA is typically measured and reported at the netting set level for a given firm. \par
U.S. regulatory authorities, the Federal Reserve and Office of the Comptroller of the Currency (OCC), periodically assess national banks' compliance with Market Risk Capital Rule (MRR). Counterparty credit risk (CCR) and funding risk (FR) metrics are key metrics used to evaluate bank risk profiles and balance sheet exposures due to over the counter (OTC) derivatives, securities financing transactions, and other transactions and exposures \citep{OCC_1130}. Basel Committee on Banking Supervision has issued supervisory guidance, in the form of its Basel III framework (and supplemental guidance), to quantify capital charges due to CCR. A new element in Basel III was a capital charge due to degradation in CCR for a given portfolio or book of business \citep{BaselCVAFramework_0715}. Potential revisions to the Basel framework may include elements to quantify CCR capital charges due to deterioration in market risk exposure.\par
The Dodd-Frank Wall Street Reform and Consumer Protection Act (July 2010) enacted regulations for the swaps market and authorized creation of centralized exchanges for swaps (and other) derivatives trading. Derivatives that trade on an exchange reference the exchange as the transaction counterparty. Since exchanges clear multiple (typically offsetting) transactions and hedge their risk through other third parties, exchange traded derivatives have minimal CCR risk profile. However, OTC derivatives typically have banks or other financial institutions as counterparties which do have material credit risk profiles. According to International Swap Dealers Association (ISDA) the OTC derivatives notional outstanding was 544 trillion at year end 2018. Interest rate derivatives notional outstanding was 437 trillion at year end 2018. Recent (04/20/20) Bloomberg CDX investment grade and high yield credit spreads are 93 and 643 basis points respectively. Consequently the CCR and FR exposures (due to uncollateralized or partially collateralized hedges) inherent in the OTC derivatives market represent significant market risk exposures. This motivates the concepts of worst case CVA, FVA, and wrong way risk (WWR) and the impact of uncertainty in probability distribution on these exposures and risk metrics. It is these considerations that motivate this line of research \citep{Ramzi19}, \citep{Hajjaji15}. \par
In our study, distributional uncertainty is characterized via the Wasserstein metric for a couple reasons. The Wasserstein metric is a (reasonably) well understood metric and a natural, intuitive way to compare two probability distributions using ideas of transport cost. It is also a flexible approach that encompasses parametric and non-parametric distributions of either discrete or continuous form. For example, one can explore distributions that alter the shape of the marginals as well as the correlation structure. Furthermore, recent duality results and structural results on the worst case distributions could help us understand and/or quantify the market model transitions as well as measure (in a relative sense) the degree of wrong way risk inherent to a given market model. \par
An outline of this paper is as follows. Section 1 gives an overview of CVA, FVA, and WWR as well as a literature review. Section 2 develops the main theoretical results of the paper and provides proof sketches. Section 3 conducts a computational study of WWR for a representative set of derivative portfolios and market environments. Section 4 discusses the conclusions and suggestions for future research. All detailed proofs of propositions, corollaries, and theorems are deferred to the Appendix. \par

\subsection{Literature Review}
\begin{remark}
The authors are not aware of any substantial research that has been done on the topic of worst case FVA. The discussion below pertains to literature regarding worst case CVA.
\end{remark}
In the past few years some research has been done to investigate and quantify the effect of distributional uncertainty on CVA. 
\citet{brigo2013counterparty} explicitly incorporate correlation into the stochastic processes driving the market risk and credit default factors. They quantify the effect of dependency structure (and hence wrong way risk) on CVA for a variety of asset classes: interest rate swaps, interest rate swaptions, commodities, equities, and foreign exchange products. 
\citet{glasserman15} bound the effect of wrong way risk on CVA. Their approach considers a discrete setting for portfolio exposures and counterparty default times and formulates worst case CVA as the solution to a worst case linear program subject to certain constraints (such as fixed marginals for portfolio exposures and default times), where the dependency structure across the risk factors is allowed to vary. As this approach leads to large values for worst case CVA, they introduce a penalty term to modulate or temper the degree of wrong way risk and run some sensitivity analysis to study the effect of the penalty term. Kullback-Leibler (KL) divergence is used to measure the distance between the reference (empirical) and the perturbed distribution. They remark that determining a suitable value for the penalty term would be a topic for further research.
\par

Memartoluie, in his PhD thesis, uses an ordered scenario copula methodology to quantify worst case CVA \citep{Amir17}. A particular method of scenario ordering correlates portfolio exposures to company default times (firm, counterparty, or both) and the resulting dependency structure introduces wrong way risk. He chooses to order exposure scenarios by increasing time averaged total exposure and then simulates company default times conditional on the exposure path using pre-specified correlation between the market risk factor(s) and credit risk factor(s). For worst case correlations set to one, he finds results for worst case CVA that are comparable to the method of \citet{glasserman15}. In a recent paper, Ben-Abdallah et al. perform a computational study on the effect of wrong way risk on CVA for a portfolio of interest rate swaps, caps, and floors \citep{Ramzi19}. They find that the dependency structure between interest rates and default intensity produces material wrong way risk whereas the dependency structure between interest rate volatility and default intensity does not.
\par

Recent results in Lagrangian duality were independently developed by \citet{blanchetFirst} and \citet{Gao16}. These results hold under mild assumptions such as upper semicontinuity in the objective function and lower semicontinuity in the distance metric. \citet{blanchet2016robust} applied this duality theory to study a number of classical regression problems in machine learning under distributional uncertainty. In that context, the authors find that distributional uncertainty can be viewed as adding a regularization term to the objective function, analogous to a penalized regression setting. Similarly, \citet{gao2017wasserstein} apply the Lagrangian duality theory to problems in statistical learning.\par
The main innovation in our work is to apply these recent results in Lagrangian duality to worst case CVA and FVA using Wasserstein distance as the ambiguity measure. Furthermore, analytical expressions are derived for the solutions to the inner and outer convex optimization problems that comprise worst case CVA and FVA via the Wasserstein approach. A computational study shows the material impact of distributional uncertainty on worst case CVA and FVA, illustrates the risk profiles, and computes the worst case distributions.\par

\subsubsection{Restatement of Lagrangian Duality Result}
In Section 2 we formulate the primal optimization problems for distributionally robust CVA and FVA. As in our earlier work this year, \citep{singh2020robust} a key step in the approach is to use recent Lagrangian duality results to formulate the equivalent dual problems. The dual problems are much more tractable than the primal problems since they only involve the reference probability measure as opposed to a Wasserstein ball of probability measures (of some finite radius). For real valued upper semicontinuous objective function $f \in L^1$ and non-negative lower semicontinuous cost function $c$ such that $\{ (u,v) : c(u,v) < \infty \}$ is Borel measurable and non-empty, it holds that \citep{BlanchetML}
\[ \sup_{ Q \in \mathcal{U}_{\delta}(Q_N) } \mathbb{E}^Q[ f(X) ] = \inf_{ \lambda \geq 0 } \: [ \lambda \delta + \frac{1}{N} \sum_{i=1}^n \Psi_\lambda(x_i) ] \]
where
\[ \Psi_\lambda(x_i) := \sup_{ u \in \text{dom} (f) } [ f(u) - \lambda c(u,x_i) ]. \]
Further details, including proofs and concrete examples, can be found in the papers by \citet{blanchetFirst}, \citet{Gao16}, and \citet{ Esfahani17}. These authors independently derived these results around the same time although \citet{blanchetFirst} did so in a more general setting. 

\subsubsection{Characterization of Worst Case Distributions}
Simply put, the set of worst case distributions (when non-empty) can be defined as $WC(f,\delta) := \{ Q^* : \mathbb{E}^{Q^*} [f(X)] = \sup_{ Q \in \mathcal{U}_{\delta}(Q_N) } \mathbb{E}^Q[ f(X) ] \}$. Another recent set of results from the literature describes the structure of the worst case distribution(s) when they exist [\citep{blanchetFirst}, \citep{Gao16}, \citep{ Esfahani17}].  The boundedness conditions for existence are tied to the growth rate $\kappa := \limsup\limits_{d(X,X_0) \rightarrow \infty} \frac{f(X) - f(X_0)}{d(X,X_0)}$ for fixed $X_0$ and the value of the dual minimizer $\lambda^*$. For empirical reference distributions, supported on $N$ points, such that $WC(f,\delta)$ is non-empty, there exists \textit{a} worst case distribution that is \textit{another} empirical distribution supported on at most $N+1$ points. This worst case distribution can be constructed via a greedy approach. For up to $N$ points, they can be identified as solving $x_i^* \in \argmin_{\tilde{x} \in dom(f)} [ \lambda^* c(\tilde{x},x_i) - f(\tilde{x}) ]$. At most one point has its probability mass split into two pieces (according to budget constraint $\delta$) that solve $x_{i_0}^{*},x_{i_0}^{**}  \in \argmin_{\tilde{x} \in dom(f)} [ \lambda^* c(\tilde{x},x_{i_0}) - f(\tilde{x}) ]$. Details can be found in \citet{Gao16}. 

\subsection{Notation and Definitions}
\subsubsection{Bilateral CVA}
Notation and core definitions for bilateral CVA (BCVA) problem setup incorporate those for unilateral CVA and DVA.
Bilateral CVA measures expected portfolio loss (or benefit) due to counterparty and/or firm default. Let $V^{+}(\tau_C)$ denote the discounted positive portfolio exposure at time $\tau_C$ and let $R_C \in [0,1)$ denote the recovery rate the firm receives upon counterparty default.  Let $V^{-}(\tau_F)$ denote the discounted negative portfolio exposure at time $\tau$ and let $R_F \in [0,1)$ denote the recovery rate the counterparty receives upon firm default. The problem setup here assumes a fixed set of observation dates, $0 = t_0 < t_1 < \cdots < t_n = T$. Let $X^+$ denote the vector of recovery adjusted discounted positive exposures and $Y_C$ denote the vector of counterparty default indicators. Let $(x^+_i,y^c_i)$ denote realizations of $(X^+,Y_C)$ along sample paths for $i = \{1,2, \ldots ,N\}$.
Let $X^-$ denote the vector of recovery adjusted discounted firm negative exposures and $Y_F$ denote the vector of firm default indicators. Let $(x^-_i,y^f_i)$ denote realizations of $(X^-,Y_F)$ along sample paths for $i = \{1,2, \ldots ,N\}$.\par
Due to the linkage, one can write $X = X^{+} + X^-$ and decompose sample realizations of $X$ accordingly.
Therefore, let $(x_i,y^c_i,y^f_i)$ denote realizations of $(X,Y_C,Y_F)$ along sample paths for  $i = \{1,2, \ldots ,N\}$. 
The relation $x_i = x^+_i + x^-_i$ can be used to decompose $x_i$ into its positive and negative exposures respectively. \par
The bilateral CVA associated with discounted positive exposure $V^{+}(\tau_C)$, counterparty default indicator $\mathbbm{1}_{\{\tau_C \leq T\} \cap \{\tau_C < \tau_F\}}$, 
discounted negative exposure $V^{-}(\tau_F)$, firm default indicator $\mathbbm{1}_{\{\tau_F \leq T\} \cap \{\tau_F < \tau_C\}}$, is
\begin{equation*}
\text{CVA}\textsuperscript{B} = \mathbb{E}[(1-R_C)V^{+}(\tau_C)\mathbbm{1}_{\{\tau_C \leq T\} \cap \{\tau_C < \tau_F\} }] + \mathbb{E}[(1-R_F)V^{-}(\tau_F)\mathbbm{1}_{\{\tau_F\leq T\} \cap \{\tau_F < \tau_C\} }].
\end{equation*}
Equivalently, one can write
\begin{equation*}
\text{CVA}\textsuperscript{B} = (1-R_C) \int_0^T \mathbb{E}[ V^+(t) | \tau_C = t, \tau_F > t] d\Pi'_C(t) + (1-R_F) \int_0^T \mathbb{E}[ V^-(t) | \tau_F = t, \tau_C > t ] d\Pi'_F(t),
\end{equation*}
where the joint counterparty and firm default time distributions are given by 
$\Pi'_C(t) = P(\tau_C \leq t, \tau_F > \tau_C)$ and $\Pi'_F(t) = P(\tau_F \leq t, \tau_C > \tau_F) \:\:$ [\citep{green2015xva}, \citep{Lichters16}, \citep{Amir17}].
The pair of vectors $(X^+,Y_C) \in (\mathbb{R}^n_{+} \times B^1_n)$ is
\begin{equation*}
X^+ = ((1-R_C)V^{+}(t_1),\ldots,(1-R_C)V^{+}(t_n)) \quad \text{and} \quad Y_C = (\mathbbm{1}_{\{\tau_C = t_1\} \cap \{\tau_F > \tau_C\}},\dots,\mathbbm{1}_{\{\tau_C = t_n\} \cap \{\tau_F > \tau_C\}}),
\end{equation*}
and the pair of vectors $(X^-,Y_F) \in (\mathbb{R}^n_{-} \times B^1_n)$ is
\begin{equation*}
X^- = ((1-R_F)V^{-}(t_1),\ldots,(1-R_F)V^{-}(t_n)) \quad \text{and} \quad Y_F = (\mathbbm{1}_{\{\tau_F = t_1\} \cap \{\tau_C > \tau_F\}},\dots,\mathbbm{1}_{\{\tau_F = t_n\} \cap \{\tau_C > \tau_F\}}).
\end{equation*}
Here $B^1_n$ denotes the set of default time vectors: binary vectors of ones and zeros with $n$ components, and at most one non-zero element. Note that counterparty or firm default occurs on at most one observation date within the fixed set of dates in the problem setup.
The empirical measure, $\Phi_N$, is
\begin{equation*}
\Phi_N(dz) = \frac{1}{N} \sum_{i=1}^{N} \mathbbm{1}_{(x_i,y^c_i,y^f_i)} (dz).
\end{equation*}
Under the empirical measure $\Phi_N$, bilateral CVA is a sum of expectations of inner products
\begin{equation*}
\text{CVA}\textsuperscript{B} = \mathbb{E}^{\Phi_N} [\langle X^+,Y_C \rangle] +  \mathbb{E}^{\Phi_N} [\langle X^-,Y_F \rangle].
\end{equation*}
In the context of this work, the uncertainty set for probability measures is
\begin{equation*}
\mathcal{U}_{\delta_3}(\Phi_N) = \{P: D_c(\Phi,\Phi_N) \leq \delta_3\}
\end{equation*}
where $D_c$ is the optimal transport cost or Wasserstein discrepancy for cost function $c$ \citep{blanchetMV}.
For convenience the definition for $D_c$ is given as
\begin{equation*}
D_c(\Phi,\Phi') = \inf \{ \mathbb{E}^\pi[c(A,B)]: \pi \in \mathcal{P}(\mathbb{R}^d \times \mathbb{R}^d), \pi_A = \Phi, \pi_B = \Phi' \}
\end{equation*}
where $\mathcal{P}$ denotes the space of Borel probability measures and $\pi_A$ and $\pi_B$ denote the distributions of $A$ and $B$. 
Here $A$ denotes $(X_A,Y^C_A,Y^F_A) \in (\mathbb{R}^n \times B^1_n \times B^1_n)$ and $B$ denotes $(X_B,Y^C_B,Y^F_B) \in (\mathbb{R}^n \times B^1_n \times B^1_n)$ respectively.
The analysis in this work uses the cost function $c_{S_3}$ where
\begin{equation*}
c_{S_3}((u,v_1,v_2),(x,y_1,y_2)) = S_3 \langle v_1-y_1,v_1-y_1 \rangle + S_3 \langle v_2-y_2,v_2-y_2 \rangle + \langle u-x,u-x \rangle.
\end{equation*}
The scale factor $S_3 > 0$ is used to compensate for different domains: $(u,v_1,v_2) \in (\mathbb{R}^n \times B^1_n \times B^1_n), (x,y_1,y_2) \in (\mathbb{R}^n \times B^1_n \times B^1_n)$.

\subsubsection{Unilateral CVA, DVA}
Bilateral CVA can be reduced to express unilateral CVA as
\begin{equation*}
\text{CVA}\textsuperscript{U} = \mathbb{E}[(1-R_C)V^{+}(\tau_C)\mathbbm{1}_{\{\tau_C \leq T\}}] = (1-R_C) \int_0^T \mathbb{E}[V^+(t) | \tau_C = t] d\Pi_C(t),
\end{equation*}
where the counterparty default time distribution is given by $\Pi_C(t) = P(\tau_C \leq t). \:\:$ 
Note the assumption here is that $\tau_C < \tau_F$. Similarly, it can be reduced to express unilateral DVA (note the minus sign), assuming $\tau_F < \tau_C$, as
\begin{equation*}
\text{DVA}\textsuperscript{U} = - \mathbb{E}[(1-R_F)V^{-}(\tau_F)\mathbbm{1}_{\{\tau_F \leq T\}}] = -(1-R_F) \int_0^T \mathbb{E}[ V^-(t) | \tau_F = t ] d\Pi_F(t),
\end{equation*}
where firm default time distribution is given by $\Pi_F(t) = P(\tau_F \leq t) \:\:$ [\citep{green2015xva}, \citep{Lichters16}, \citep{Amir17}].

\subsubsection{FVA}
Notation and core definitions for (bilateral) FVA problem setup incorporate those for FCA and FBA.
FVA measures expected funding costs and benefits over portfolio lifetime. Let $V^{+}(t)$ denote the positive portfolio exposure at time $t$.  Let $V^{-}(t)$ denote the negative portfolio exposure at time $t$. The problem setup here assumes a fixed set of observation dates, $0 = t_0 < t_1 < \cdots < t_n = T$.
Let $X^+$ denote the vector of discounted positive exposures and $Y_C$ denote the vector of counterparty survival indicators. 
Let $X^-$ denote the vector of discounted negative exposures and $Y_F$ denote the vector of firm survival indicators. 
Further, let $Y_{CF}$ denote the Hadamard product $Y_C \odot Y_F$ which represents the vector of joint survival indicators. To incorporate funding, let $Z^+$ denote the vector of funding costs incurred on exposures $X^+$. And similarly for $Z^-$ with respect to exposures $X^-$.
Due to the linkage between $Z^+$ and $Z^-$, one can write $Z = Z^{+} + Z^-$ and decompose sample realizations of $Z$ into $Z^+$ and $Z^-$ accordingly.
Therefore, let $(z_i,y^{cf}_i)$ denote realizations of $(Z,Y_{CF})$ along sample paths for  $i = \{1,2, \ldots ,N\}$. 
The relation $z_i = z^+_i + z^-_i$ can be used to decompose $z_i$ into its positive and negative exposures respectively. \par


The FVA associated with funding costs $Z(t)$, joint survival indicator $\mathbbm{1}_{\{\tau_C > t\} \cap \{\tau_F > t\}}$ is [\citep{Lichters16}, \citep{green2015xva}]:
\begin{equation*}
\text{FVA} = \text{FCA} + \text{FBA} = \int_0^T \mathbb{E}[Z^+(t) \mathbbm{1}_{\{\tau_C > t\} \cap \{\tau_F > t\}} ] dt + \int_0^T \mathbb{E}[Z^-(t) \mathbbm{1}_{\{\tau_C > t\} \cap \{\tau_F > t\}} ] dt =  \int_0^T \mathbb{E}[Z(t) \mathbbm{1}_{\{\tau_C > t\} \cap \{\tau_F > t\}} ] dt.
\end{equation*}
\noindent The pair of vectors $(Z,Y_{CF}) \in (\mathbb{R}^n \times B^1_n)$ is
\begin{equation*}
Z = (Z^{+}(t_1)+Z^{-}(t_1),\ldots,Z^{+}(t_n)+Z^{-}(t_n)) \quad \text{and} \quad Y_{CF} = (\mathbbm{1}_{\{\tau_C > t_1\} \cap \{\tau_F > t_1\}},\dots,\mathbbm{1}_{\{\tau_C > t_n\} \cap \{\tau_F > t_n\}}),
\end{equation*}
and the pair of vectors $(Z^+,Z^-) \in (\mathbb{R}^n_{+} \times \mathbb{R}^n_{-})$ is
\begin{equation*}
Z^+ = (f_c(t_0,t_1)X^{+}(t_1),\ldots,f_c(t_{n-1},t_n)X^{+}(t_n)) \quad \text{and} \quad Z^- = (f_b(t_0,t_1)X^{-}(t_1),\ldots,f_b(t_{n-1},t_n)X^{-}(t_n)).
\end{equation*}
Here $B^1_n$ denotes the set of survival time vectors: binary vectors of ones and zeros with $n$ components, and at most one block of ones followed by a complementary block of zeros.
The empirical measure, $\Phi_N$, is
\begin{equation*}
\Phi_N(dz) = \frac{1}{N} \sum_{i=1}^{N} \mathbbm{1}_{(z_i,y^{cf}_i)} (dz).
\end{equation*}
Under the empirical measure $\Phi_N$, FVA is a sum of expectations of inner products
\begin{equation*}
\text{FVA} = \mathbb{E}^{\Phi_N} [\langle Z^+,Y_{CF} \rangle] +  \mathbb{E}^{\Phi_N} [\langle Z^-,Y_{CF} \rangle] = \mathbb{E}^{\Phi_N} [\langle Z,Y_{CF} \rangle].
\end{equation*}
In the context of this work, the uncertainty set for probability measures is
\begin{equation*}
\mathcal{U}_{\delta_3}(\Phi_N) = \{P: D_c(\Phi,\Phi_N) \leq \delta_3\}
\end{equation*}
where $D_c$ is the optimal transport cost or Wasserstein discrepancy for cost function $c$ \citep{blanchetMV}.
For convenience the definition for $D_c$ is
\begin{equation*}
D_c(\Phi,\Phi') = \inf \{ \mathbb{E}^\pi[c(A,B)]: \pi \in \mathcal{P}(\mathbb{R}^d \times \mathbb{R}^d), \pi_A = \Phi, \pi_B = \Phi' \}
\end{equation*}
where $\mathcal{P}$ denotes the space of Borel probability measures and $\pi_A$ and $\pi_B$ denote the distributions of $A$ and $B$. 
Here $A$ denotes $(Z_A,Y_A) \in (\mathbb{R}^n \times B^1_n)$ and $B$ denotes $(Z_B,Y_B) \in (\mathbb{R}^n \times B^1_n)$ respectively.
This work uses the cost function $c_{S_3}$ where
\begin{equation*}
c_{S_3}((u,v),(z,y)) = S_3 \langle v-y,v-y \rangle + \langle u-z,u-z \rangle.
\end{equation*}
The scale factor $S_3 > 0$ is used to compensate for different domains: $(u,v) \in (\mathbb{R}^n \times B^1_n), (z,y) \in (\mathbb{R}^n \times B^1_n)$. 

\begingroup
\setlength{\parindent}{0pt}
\section{Theory: Robust XVA and Wrong Way Risk}
\subsection{Unilateral CVA, DVA}
The robust unilateral CVA can be written as
\begin{equation*}\label{eqn:primal}
\sup_{P \in \mathcal{U}_{\delta_1}(P_N)} \mathbb{E}^P [\langle X^+, Y_C \rangle]  \tag{P1}.
\end{equation*}
Similarly, the robust unilateral DVA is
\begin{equation*}\label{eqn:primal2}
- \sup_{Q \in \mathcal{U}_{\delta_2}(Q_N)} \mathbb{E}^Q [\langle X^-, Y_F \rangle]  \tag{P2}.
\end{equation*}
As such, the dual formulations and solutions to the above primal optimization problems are special cases of the solutions to the bilateral CVA optimization problems, to be described next.
\subsection{Bilateral CVA}
\subsubsection{Inner Optimization Problem}
The robust bilateral CVA is
\begin{equation*}\label{eqn:primal3}
\sup_{\Phi \in \mathcal{U}_{\delta_3}(\Phi_N)} \mathbb{E}^{\Phi} [ \langle X^+, Y_C \rangle + \langle X^-, Y_F \rangle] \tag{P3}.
\end{equation*}

Similar to before, use recent duality results, noting that the inner product $\langle \: ; \rangle$ satisfies the upper semicontinuous condition of the Lagrangian duality theorem, and cost function $c_S$ satisfies the non-negative lower semicontinuous condition (see \citet{blanchetFirst} Assumptions 1 \& 2, \citet{Gao16}). Hence the dual problem can be written as
\begin{equation*}\label{eqn:dual3}
 \inf_{\alpha \geq 0} \:  F(\alpha) := \bigg[ \alpha \delta_3+ \frac{1}{N} \sum_{i=1}^{N} \Psi_{\alpha}(x_i,y^c_i,y^f_i) \bigg]  \tag{D3}
\end{equation*}
where 
\begin{align*}
\Psi_\alpha(x_i,y^c_i,y^f_i)  = \sup_{u \in \mathbb{R}^n, v_1 \in {B^1_n}, v_2 \in {B^1_n}} &[  \langle u^+, \mathbbm{1}_{\{v_1 < v_2\}} v_1 \rangle + \langle u^-, \mathbbm{1}_{\{v_2 < v_1\}} v_2 \rangle - \alpha c_{S_3}((u,v_1,v_2),(x_i,y^c_i,y^f_i)) ] \\
								  = \sup_{u \in \mathbb{R}^n, v_1 \in {B^1_n},v_2 \in {B^1_n}} &[ \langle u^+, \mathbbm{1}_{\{v_1 < v_2\}} v_1 \rangle + \langle u^-, \mathbbm{1}_{\{v_2 < v_1\}} v_2 \rangle  - \alpha( \langle u-x_i, u-x_i \rangle + S_3 \langle v_1-y^c_i, v_1-y^c_i \rangle \\
								  &+ S_3 \langle v_2-y^f_i, v_2-y^f_i \rangle )  ].
\end{align*}


Note that default times $(v_1,v_2)$ are compared via the indicator function $\mathbbm{1}_{\{ v_1 \lessgtr v_2 \}}$ by comparing indices (into the fixed dates array $0 < t_1 < \cdots < t_n = T$) of the respective default times.
So if $v_1$ has a one element in index $i$ and either $\|v_2\| = 0$ or $v_2$ has a one element in index $j$ and $i < j$ then $\mathbbm{1}_{\{ v_1 < v_2 \}} = 1$ else if $i > j$ or $\|v_1\| = 0$ then $\mathbbm{1}_{\{ v_1 < v_2 \}} = 0$. The probability that $i = j \: $ for any $i,j \in \{1,\dots,n\}$ is zero in continuous time, hence this case is not considered here.
Also $\|v_1\| = 1$ implies default time $v_1 \leq t_n = T$, the maturity date of the CVA calculation. Similar analysis applies to $v_2$.\\

Now apply change of variables $w_1 = (u-x_i)$, $w_2 = (v_1-y^c_i)$, and $w_3 = (v_2-y^f_i)$ to get
\begin{align*}
\begin{split}
\Psi_\alpha(x_i,y^c_i,y^f_i)  = \sup_{w_1 \in \mathbb{R}^n, w_2 \in {B^2_n},w_3 \in {B^2_n}} &\big[ \langle (w_1+x_i)^+, \mathbbm{1}_{\{w_2+y^c_i < w_3+y^f_i\}} w_2+y^c_i \rangle 
+  \langle (w_1+x_i)^-, \mathbbm{1}_{\{w_3+y^f_i < w_2+y^c_i\}} w_3+y^f_i \rangle \\
&- \alpha ( \langle w_1, w_1 \rangle + S_3 \langle w_2, w_2 \rangle + S_3 \langle w_3, w_3 \rangle ) \big].
\end{split}
\end{align*}

It turns out that $\Psi_\alpha$ can be expressed as the pointwise max of four functions of more complex forms.
The four functions represent the four logical cases for $w_2$ and $w_3$ each being zero or non-zero. Furthermore, we need to consider the sub-cases where the counterparty defaults before the firm, as in $\Psi_\alpha^a$ or vice-versa as in $\Psi_\alpha^b$. Again, $\Psi_\alpha$ quantifies the adversarial moves in CVA and DVA across both time and spatial dimensions while accounting for the associated cost via the $K$ terms. 
\begin{remark}
Note that this result involves some lengthy and tedious derivations and requires some time to go through. However, there are some patterns across the various cases and sub-cases which does simplify the analysis to some extent. 
\end{remark}
\renewcommand{\arraystretch}{1.75}
\begin{table}[h]
\normalsize
\begin{center}
\caption{Lookup table of optimization sub-problems}
\begin{tabular}{ |c|c|l| }
 \hline
\textit{Optimization} & \textit{Objective Function} & \textit{Solution} \\
 \hline
$\sup_{w_1 \in \mathbb{R}^n}$ & $\langle w_1, y_i \rangle - \alpha \langle w_1, w_1 \rangle$ & $\frac{\| y_i \|^2}{4 \alpha}$ \\ 
\hline
$\sup_{w_1 \leq x_{i\tau_2}}$ & $\langle w_1, y_i \rangle - \alpha \langle w_1, w_1 \rangle$ & $ [x_{i\tau_2} \wedge \frac{\|y_i\|}{2 \alpha}] - \alpha [x_{i\tau_2} \wedge \frac{\|y_i\|}{2 \alpha}]^2 $ \\ 
 \hline
$\sup_{w_1 \in \mathbb{R}^n}$ & $\langle (w_1 + x_i)^+, y_i \rangle - \alpha \langle w_1, w_1 \rangle$ & $[ \frac{1}{4 \alpha} + \langle x_i, y_i \rangle]^+$ \\ 
\hline
$\sup_{w_1 \in \mathbb{R}^n}$ & $\langle (w_1 + x_i)^-, y_i \rangle - \alpha \langle w_1, w_1 \rangle$ & $\mathbbm{1}_{\{(x_{i\tau_2}<-\frac{1}{2\alpha}) \vee (x_{i\tau_2}>0)\}} \big[\frac{1}{4\alpha} + \langle x_i, y_i \rangle \big]^-  -
      \mathbbm{1}_{\{-\frac{1}{2\alpha} \leq x_{i\tau_2} \leq 0\}} \big[ \alpha (\langle x_i, y_i \rangle)^2 \big]$ \\ 
\hline
$\sup_{w_1 \in \mathbb{R}^n}$ & $(w_1 + x_{i\tau_1})^- - \alpha \langle w_1, w_1 \rangle$ & $\mathbbm{1}_{\{(x_{i\tau_1}<-\frac{1}{2\alpha}) \vee (x_{i\tau_1}>0)\}} \big[\frac{1}{4\alpha} +  \langle x_i, y_i \rangle + (x_{i\tau_1} - x_{i\tau_2}) \big]^-$ \\  
& & $-\mathbbm{1}_{\{-\frac{1}{2\alpha} \leq x_{i\tau_1} \leq 0\}} \big[ \alpha ( x_{i\tau_1})^2 \big]$ \\ 
\hline
\end{tabular} 
\end{center} 
\end{table}
\renewcommand{\arraystretch}{1}

\vspace{0.2cm}
\begin{proprep}
We have \, $\Psi_\alpha(x_i,y^c_i,y^f_i) = \bigvee_{k=1}^4  \Psi_\alpha^{k}(x_i,y^c_i,y^f_i)$ \, where\\
$\Psi_\alpha^1(x_i,y^c_i,y^f_i)  = \mathbbm{1}_{(y_i^c < y_i^f)} \Psi_\alpha^{1a}(x_i,y^c_i,y^f_i) + \mathbbm{1}_{(y_i^f < y_i^c)} \Psi_\alpha^{1b}(x_i,y^c_i,y^f_i)$,\\
$\Psi_\alpha^{2}(x_i,y^c_i,y^f_i)  = \mathbbm{1}_{(w_2+y_i^c < y_i^f)} \Psi_\alpha^{2a}(x_i,y^c_i,y^f_i) + \mathbbm{1}_{(y_i^f < w_2+ y_i^c)} \Psi_\alpha^{2b}(x_i,y^c_i,y^f_i)$,\\
$\Psi_\alpha^{3}(x_i,y^c_i,y^f_i)  = \mathbbm{1}_{(y_i^c < w_3+y_i^f)} \Psi_\alpha^{3a}(x_i,y^c_i,y^f_i) + \mathbbm{1}_{(w_3+y_i^f < y_i^c)} \Psi_\alpha^{3b}(x_i,y^c_i,y^f_i)$,\\
$\Psi_\alpha^{4}(x_i,y^c_i,y^f_i)  = \mathbbm{1}_{(w_2+y_i^c < w_3+y_i^f)} \Psi_\alpha^{4a}(x_i,y^c_i,y^f_i) + \mathbbm{1}_{(w_3+y_i^f < w_2+y_i^c)} \Psi_\alpha^{4b}(x_i,y^c_i,y^f_i)$,\\ \\
and (suppressing arguments for brevity):\\
$\Psi_\alpha^{1a} = \bigg[\frac{1}{4\alpha} + \langle x_i, y_i^c \rangle \bigg]^+$,
$\Psi_\alpha^{1b} = \bigg[ \mathbbm{1}_{\{(x_{i\tau_2}<-\frac{1}{2\alpha}) \vee (x_{i\tau_2}>0)\}} \big[\frac{1}{4\alpha} + \langle x_i, y_i^f \rangle \big]^-  -
      \mathbbm{1}_{\{-\frac{1}{2\alpha} \leq x_{i\tau_2} \leq 0\}} \big[ \alpha (\langle x_i, y_i^f \rangle)^2 \big] \bigg]$,\\
$\Psi^{2a} = \bigg[ \big[ \frac{1}{4 \alpha} + \langle x_i,y_i^c \rangle + ( x_{i\tau_1^*} - x_{i\tau_2} ) \big]^+ - \alpha S_3 K^{2a} \bigg]$,
$\Psi_\alpha^{2b} = \bigg[ \Psi_\alpha^{1b} - \alpha S_3 K^{2b} \bigg]$,\\
$\Psi^{3a} = \bigg[ \Psi_\alpha^{1a} - \alpha S_3 K^{3a} \bigg]$,
$\Psi^{3b} =  \bigg[ \mathbbm{1}_{\{(x_{i\tau_1^*}<-\frac{1}{2\alpha}) \vee (x_{i\tau_1^*}>0)\}} \big[\frac{1}{4\alpha} + \langle x_i, y_i^f \rangle + (x_{i\tau_1^*} - x_{i\tau_2}) \big]^-  -
      \mathbbm{1}_{\{-\frac{1}{2\alpha} \leq x_{i\tau_1^*} \leq 0\}} \big[ \alpha (x_{i\tau_1^*})^2 \big] - \alpha S_3 K^{3b} \bigg]$,\\
$\Psi^{4a} =  \bigg[ \Psi^{2a} - \alpha S_3 ( K^{4a} - K^{2a} ) \bigg]$,
$\Psi^{4b} =  \bigg[ \Psi^{3b} - \alpha S_3 ( K^{4b} - K^{3b} ) \bigg]$.\\
Note parameter $\tau_1^*$ and constant $K$ are defined within the proof by $\mathbf{cases}$ (see Supplementary Material), and are omitted here for brevity.
Recall $\tau_2$ is index $\tau$ such that $y_{i\tau}^{\{c,f\}} = 1$ else it is 0 if $\| y_i^{\{c,f\}} \| = 0$. The selection in $\{c,f\}$ is determined by context.
\end{proprep}
\begin{proofsketch}
This result follows from jointly maximizing the adversarial exposure $w_1$ and the default time indices $w_2, w_3$. The structure of $B^2_n$ allows us to decouple this joint maximization and find the critical point to maximize the quadratic in $w_1$ and write down the condition to select the optimal default time index $\tau_1^*$ for either the counterparty (in sub-case a) or the firm (in sub-case b), as determined by first to default. Finally, take the max over the four logical cases for $w_2$ and $w_3$ to arrive at the solution. The $K$ terms represent the cost associated with the worst case BCVA.
\end{proofsketch}

\begin{appendixproof}
\begingroup
\setlength{\parindent}{0pt}

The particular structure of $B^1_n$ and $B^2_n$ will be exploited to evaluate the $\sup$ above.
The analysis proceeds by considering different cases for optimal values $(w_1^{*}, w_2^{*},w_3^{*})$. \\ \\
$
\mathbf{Case \, 1}
$
\quad Suppose $w_2^{*} = 0, w_3^{*} = 0$. Then\\
\begin{equation*}
\Psi_\alpha(x_i,y^c_i,y^f_i)  = \sup_{w_1 \in \mathbb{R}^n} \big[ \langle (w_1+x_i)^+, \mathbbm{1}_{\{y^c_i < y^f_i\}} y^c_i \rangle 
+  \langle (w_1+x_i)^-, \mathbbm{1}_{\{y^f_i < y^c_i\}} y^f_i \rangle 
- \alpha ( \langle w_1, w_1 \rangle ) \big].
\end{equation*}

\begin{enumerate}[label*=\alph*)]
\item Suppose $\mathbbm{1}_{(y_i^c < y_i^f)} = 1$. Then
\begin{equation*}
\Psi_\alpha(x_i,y^c_i,y^f_i)  = \sup_{w_1 \in \mathbb{R}^n} \big[ \langle (w_1+x_i)^+, y^c_i \rangle - \alpha ( \langle w_1, w_1 \rangle ) \big].
\end{equation*}
Therefore $\|y_i^c\| = 1$. Let $\tau_2$ denote default time for $y^c_i$. Simplify further to get
\begin{equation*}
\Psi_\alpha(x_i,y^c_i,y^f_i)  = \sup_{w_{1\tau_2} \in \mathbb{R}} \big[ (w_{1\tau_2}+x_{i\tau_2})^+ - \alpha ( w_{1\tau_2} )^2 \big].
\end{equation*}
Now follow the approach in \cite {Bartl17} to write down the first order optimality condition:
\begin{equation*}
\mathbbm{1}_{[0,\infty)} (w_{1\tau_2}+x_{i\tau_2}) - 2 \alpha w_{1\tau_2} \leq 0 \leq \mathbbm{1}_{(0,\infty)} (w_{1\tau_2}+x_{i\tau_2}) - 2 \alpha w_{1\tau_2}.
\end{equation*}
  \begin{enumerate}[label=\roman*)]
	\item Suppose $(w_{1\tau_2}^* + x_{i\tau_2}) < 0$.
	Then $w_{1\tau_2}^* = 0$. So $x_{i\tau_2} < 0 \implies w_{i\tau_2}^* = 0$.
	\item Suppose $(w_{1\tau_2}^* + x_{i\tau_2}) > 0$.
	Then $w_{1\tau_2}^* = \frac{1}{2\alpha}$. So $x_{i\tau_2} > - \frac{1}{2\alpha} \implies w_{1\tau_2}^* = \frac{1}{2\alpha}$.
	\item Note $(w_{1\tau_2}^* + x_{i\tau_2}) = 0$ is not possible (does not satisfy first order optimality condition).
	\\ \\ Considering the intervals for $x_{i\tau_2}$ above, there are three cases as below.
  \end{enumerate}

  \begin{enumerate}[label=\roman*)]
	\item $x_{i\tau_2} \geq 0 \implies w_{1\tau_2}^* = \frac{1}{2\alpha} \implies \Psi_\alpha = [\frac{1}{4\alpha} + x_{i\tau_2}]$.
	\item $x_{i\tau_2} \leq - \frac{1}{2\alpha} \implies w_{1\tau_2}^* = 0 \implies \Psi_\alpha = 0$.
	\item $(- \frac{1}{2\alpha} < x_{i\tau_2} < 0 ) \implies \Psi_\alpha = [\frac{1}{4\alpha} + x_{i\tau_2}]^+$.
  \end{enumerate}
  In summary, considering all cases above, conclude that
  \begin{equation*} 
  \Psi_\alpha^{1a}(x_i,y^c_i,y^f_i)  = \big[\frac{1}{4\alpha} + x_{i\tau_2} \big]^+ .
  \end{equation*}
  This can also be expressed as
  \begin{equation*} 
  \Psi_\alpha^{1a}(x_i,y^c_i,y^f_i)  = \big[\frac{1}{4\alpha} + \langle x_i, y_i^c \rangle \big]^+.
  \end{equation*}

\item Suppose $\mathbbm{1}_{(y_i^f < y_i^c)} = 1$. Then
\begin{equation*}
\Psi_\alpha(x_i,y^c_i,y^f_i)  = \sup_{w_1 \in \mathbb{R}^n} \big[ \langle (w_1+x_i)^-, y^f_i \rangle - \alpha ( \langle w_1, w_1 \rangle ) \big].
\end{equation*}
Therefore $\|y_i^f\| = 1$. Let $\tau_2$ denote default time for $y^f_i$. Simplify further to get
\begin{equation*}
\Psi_\alpha(x_i,y^c_i,y^f_i)  = \sup_{w_{1\tau_2} \in \mathbb{R}} \big[ (w_{1\tau_2}+x_{i\tau_2})^- - \alpha ( w_{1\tau_2} )^2 \big].
\end{equation*}
Now follow the approach in \cite {Bartl17} to write down the first order optimality condition:
\begin{equation*}
\mathbbm{1}_{(-\infty,0]} (w_{1\tau_2}+x_{i\tau_2}) - 2 \alpha w_{1\tau_2} \leq 0 \leq \mathbbm{1}_{(-\infty,0)} (w_{1\tau_2}+x_{i\tau_2}) - 2 \alpha w_{1\tau_2}.
\end{equation*}
  \begin{enumerate}[label=\roman*)]
	\item Suppose $(w_{1\tau_2}^* + x_{i\tau_2}) > 0$.
	Then $w_{1\tau_2}^* = 0$. So $x_{i\tau_2} > 0 \implies w_{1\tau_2}^* = 0$.
	\item Suppose $(w_{1\tau_2}^* + x_{i\tau_2}) < 0$.
	Then $w_{1\tau_2}^* = \frac{1}{2\alpha}$. So $x_{i\tau_2} < - \frac{1}{2\alpha} \implies w_{1\tau_2}^* = \frac{1}{2\alpha}$.
	\item Note $(w_{1\tau_2}^* + x_{i\tau_2}) = 0$ is not possible (does not satisfy first order optimality condition).
	\\ \\ Considering the intervals for $x_{i\tau_2}$ above, there are three cases as below.
  \end{enumerate}

 \begin{enumerate}[label=\roman*)]
	\item $x_{i\tau_2} > 0 \implies w_{1\tau_2}^* = 0 \implies \Psi_\alpha = 0$.
	\item $x_{i\tau_2} < - \frac{1}{2\alpha} \implies w_{i\tau_2}^* = \frac{1}{2\alpha} \implies \Psi_\alpha = [\frac{1}{4\alpha} + x_{i\tau_2}]$.
	\item $[ - \frac{1}{2\alpha} \leq x_{i\tau_2} \leq 0 ] \implies w_{1\tau_2}^* = |x_{i\tau_2}|$.\\
	Note the slope $(1 - 2 \alpha w_{1\tau_2})$ is positive for $0 \leq w_{1\tau_2} < \frac{1}{2\alpha}$, and equals zero at $w_{1\tau_2} = \frac{1}{2\alpha}$.\\
	However, $(w_{1\tau_2} + x_{i\tau_2})^-$ attains its max value of zero for $w_{i\tau_2} = |x_{i\tau_2}|$ so stop there.
  \end{enumerate}
  In summary, considering all cases above, conclude that 
  \begin{equation*}
  \Psi_\alpha^{1b}(x_i,y^c_i,y^f_i) = \bigg[ \mathbbm{1}_{\{(x_{i\tau_2}<-\frac{1}{2\alpha}) \vee (x_{i\tau_2}>0)\}} \big[\frac{1}{4\alpha} + x_{i\tau_2}\big]^-  -
      \mathbbm{1}_{\{-\frac{1}{2\alpha} \leq x_{i\tau_2} \leq 0\}} \big[ \alpha (x_{i\tau_2})^2 \big] \bigg].
  \end{equation*}
  This can also be expressed as
\begin{equation*}
  \Psi_\alpha^{1b}(x_i,y^c_i,y^f_i)  = \bigg[ \mathbbm{1}_{\{(x_{i\tau_2}<-\frac{1}{2\alpha}) \vee (x_{i\tau_2}>0)\}} \big[\frac{1}{4\alpha} + \langle x_i, y_i^f \rangle \big]^-  -
      \mathbbm{1}_{\{-\frac{1}{2\alpha} \leq x_{i\tau_2} \leq 0\}} \big[ \alpha (\langle x_i, y_i^f \rangle)^2 \big] \bigg].
  \end{equation*}

\item Suppose $\mathbbm{1}_{(\|y_i^f\| = \|y_i^c\| = 0)} = 1$.\\
  In this trivial case, $\Psi_\alpha = 0$. Note there is no third subcase for $\text{Cases 2-4}$ below since that would imply $w_2^* = 0, w_3^* = 0$.
\end{enumerate}

Finally, to sum up Case 1, considering parts a) and b), let us write:
\begin{equation*}
\Psi_\alpha^1(x_i,y^c_i,y^f_i)  = \mathbbm{1}_{(y_i^c < y_i^f)} \Psi_\alpha^{1a}(x_i,y^c_i,y^f_i) + \mathbbm{1}_{(y_i^f < y_i^c)} \Psi_\alpha^{1b}(x_i,y^c_i,y^f_i).
\end{equation*}

$
\mathbf{Case \, 2}
$
\quad Suppose $w_2^{*} \neq 0, w_3^{*} = 0$.\\
Then $w_2^*$ has +1 in position $\tau_1^*$ and -1 in position $\tau_2$, where $\tau_j = 0$ means the value $\pm 1$ does not occur. \\
Furthermore, $\tau_1^* \neq \tau_2$ otherwise $w_2^* = 0$.
\begin{equation*}
\Psi_\alpha(x_i,y^c_i,y^f_i)  = \sup_{w_1 \in \mathbb{R}^n, w_2 \in B_n^2} \big[ \langle (w_1+x_i)^+, \mathbbm{1}_{\{w_2 + y^c_i < y^f_i\}} w_2 + y^c_i \rangle 
+  \langle (w_1+x_i)^-, \mathbbm{1}_{\{y^f_i < w_2 + y^c_i\}} y^f_i \rangle 
- \alpha ( \langle w_1, w_1 \rangle + S_3 \langle w_2, w_2 \rangle ) \big].
\end{equation*}

\begin{enumerate}[label*=\alph*)]
\item Suppose $\mathbbm{1}_{(w_2 + y_i^c < y_i^f)} = 1$. Then
\begin{equation*}
\Psi_\alpha(x_i,y^c_i,y^f_i)  = \sup_{w_1 \in \mathbb{R}^n, w_2 \in B_n^2} \big[ \langle (w_1+x_i)^+,  w_2 + y^c_i \rangle 
- \alpha ( \langle w_1, w_1 \rangle + S_3 \langle w_2, w_2 \rangle ) \big].
\end{equation*}
Recall $\langle (w_1 + x_i), (w_2+y_i^c) \rangle = (w_{1\tau_1} + x_{i\tau_1})$. 
Also recall $\tau_1$ and $\tau_2$ are associated with $y^c_i$.
Let $\tau_{2,f}$ denote default time (index) for $y_i^f$.
The default time constraint implies $\tau_1 < \tau_{2,f}$.
Therefore $\tau_1 > 0$.
The structure of finite set $B^2_n$ implies
\begin{equation*}
\Psi_\alpha(x_i,y^c_i,y^f_i) = \sup_{w_1 \in \mathbb{R}^n, 0 < \tau_1 < \tau_{2,f}, \tau_1 \neq \tau_2} \big[  (w_{1\tau_1} + x_{i\tau_1})^+  - \alpha ( \langle w_1, w_1 \rangle + S_3 \langle w_2, w_2 \rangle ) \big].
\end{equation*}
Observe the only positive component for $w_1 \in \mathbb{R}^n$ in $\sup$ above is $\tau_1$.\\
\begin{equation*}
\sup_{w_1 \in \mathbb{R}^n} \big[ (w_{1\tau_1}+x_{i\tau_1})^+ - \alpha \langle w_1, w_1 \rangle \big] = \sup_{w_{1\tau_1} \in \mathbb{R}} \big[ (w_{1\tau_1}+x_{i\tau_1})^+ - \alpha (w_{1\tau_1}^2 ) \big].
\end{equation*}
Evaluating at the critical point $w^{*}_{1\tau_1} = \frac{1}{2\alpha} \in \mathbb{R}$ for the above quadratic gives
\begin{equation*}
\sup_{w_{1\tau_1} \in \mathbb{R}} \big[ (w_{1\tau_1}+ x_{i\tau_1})^+  - \alpha (w_{1\tau_1}^2 ) \big] = \big[\frac{1}{4\alpha} + x_{i\tau_1}\big]^+.
\end{equation*}
Therefore one can write
\begin{equation*}
\Psi_\alpha(x_i,y^c_i,y^f_i) = \max_{0 < \tau_1 < \tau_{2,f}, \tau_1 \neq \tau_2} \big[ \frac{1}{4\alpha} + x_{i\tau_1} \big]^+  - \alpha S_3 K^{2a}
\end{equation*}
where $K^{2a} := ( 1 + \mathbbm{1}_{\{ \tau_2 \neq 0 \} } )$.
Furthermore, $\tau_1^{*}$ is determined as
\begin{equation*}
\tau_1^{*} = \argmax_{0 < \tau_1 < \tau_{2,f}, \tau_1 \neq \tau_2} [ x^+_{i\tau_1} ].
\end{equation*}
Substituting back into expression for $\Psi_\alpha$ gives
\begin{equation*}
\Psi^{2a}_\alpha(x_i,y^c_i,y^f_i) = \bigg[ \big[ \frac{1}{4 \alpha} + x_{i\tau_1^{*}} \big]^+ - \alpha S_3 K^{2a} \bigg].
\end{equation*}
This can also be expressed as
\begin{equation*}
\Psi^{2a}_\alpha(x_i,y^c_i,y^f_i) = \bigg[ \big[ \frac{1}{4 \alpha} + \langle x_i,y_i^c \rangle + ( x_{i\tau_1^*} - x_{i\tau_2} ) \big]^+ - \alpha S_3 K^{2a} \bigg].
\end{equation*}
\item Suppose $\mathbbm{1}_{(y_i^f < w_2 + y_i^c)} = 1$. Then
\begin{equation*}
\Psi_\alpha(x_i,y^c_i,y^f_i)  = \sup_{w_1 \in \mathbb{R}^n, w_2 \in B_n^2} \big[  \langle (w_1+x_i)^-, y^f_i \rangle 
- \alpha ( \langle w_1, w_1 \rangle + S_3 \langle w_2, w_2 \rangle ) \big].
\end{equation*}
Recall $\tau_1$ and $\tau_2$ are associated with $y^c_i$.
Let $\tau_{2,f}$ denote the default time (index) for $y_i^f$.
The default time constraint implies $\tau_{2,f} < \tau_1$. Therefore $\tau_{2,f} > 0$ and $\|y_i^f\| = 1$.
Note the only non-zero component of $\|y^f_i\|$ is $\tau_{2,f}$. 
Hence set $w^*_{1\tau} = 0 \: \forall \tau \neq \tau_{2,f}$. Simplifying further
\begin{equation*}
\Psi_\alpha(x_i,y^c_i,y^f_i)  = \sup_{w_{1\tau_{2,f}} \in \mathbb{R}, w_2 \in B_n^2} \big[ (w_{1\tau_{2,f}}+x_{i\tau_{2,f}})^- - \alpha ( ( w_{1\tau_{2,f}} )^2 + S_3 K^{2b} ) \big].
\end{equation*}
where $K^{2b} := ( \mathbbm{1}_{\{ \tau_1 \neq 0 \} } + \mathbbm{1}_{\{ \tau_2 \neq 0 \} } )$ = 1.
For $K^{2b}$, if $\tau_2 = 0$, then $\tau_1 \neq 0$ since $w_2^* \neq 0$. Otherwise set $\tau_1 = 0$ if $\tau_2 \neq 0$ to maximize $\sup_{w_2}$ above.
Following the calculations in $\text{Case} \: \text{1b})$ above, conclude that
\begin{equation*}
  \Psi_\alpha^{2b}(x_i,y^c_i,y^f_i)  = \bigg[ \mathbbm{1}_{\{(x_{i\tau_2}<-\frac{1}{2\alpha}) \vee (x_{i\tau_2}>0)\}} [\frac{1}{4\alpha} + x_{i\tau_2}]^-  -
      \mathbbm{1}_{\{-\frac{1}{2\alpha} \leq x_{i\tau_2} \leq 0\}} \big[ \alpha (x_{i\tau_2})^2  \big] - \alpha S_3 K^{2b} \bigg].
 \end{equation*}
This can also be expressed as
\begin{equation*}
  \Psi_\alpha^{2b}(x_i,y^c_i,y^f_i)  = \bigg[ \mathbbm{1}_{\{(x_{i\tau_2}<-\frac{1}{2\alpha}) \vee (x_{i\tau_2}>0)\}} \big[\frac{1}{4\alpha} + \langle x_i, y_i^f \rangle \big]^-  -
      \mathbbm{1}_{\{-\frac{1}{2\alpha} \leq x_{i\tau_2} \leq 0\}} \big[ \alpha ( \langle x_i, y_i^f \rangle )^2 \big] - \alpha S_3 K^{2b} \bigg].
 \end{equation*}

Finally, to sum up Case 2, considering parts a) and b), let us write:
\begin{equation*}
\Psi_\alpha^{2}(x_i,y^c_i,y^f_i)  = \mathbbm{1}_{(w_2+y_i^c < y_i^f)} \Psi_\alpha^{2a}(x_i,y^c_i,y^f_i) + \mathbbm{1}_{(y_i^f < w_2+ y_i^c)} \Psi_\alpha^{2b}(x_i,y^c_i,y^f_i).
\end{equation*}
\end{enumerate}
$
\mathbf{Case \, 3}
$
\quad Suppose $w_2^{*} = 0, w_3^{*} \neq 0$.\\
Then $w_3^*$ has +1 in position $\tau_1^*$ and -1 in position $\tau_2$, where $\tau_j = 0$ means the value $\pm 1$ does not occur. \\
Furthermore, $\tau_1^* \neq \tau_2$ otherwise $w_3^* = 0$.
\begin{equation*}
\Psi_\alpha(x_i,y^c_i,y^f_i)  = \sup_{w_1 \in \mathbb{R}^n, w_3 \in B_n^2} \big[ \langle (w_1+x_i)^+, \mathbbm{1}_{\{y^c_i < w_3+y^f_i\}} y^c_i \rangle 
+  \langle (w_1+x_i)^-, \mathbbm{1}_{\{w_3+y^f_i < y^c_i\}} w_3+y^f_i \rangle 
- \alpha ( \langle w_1, w_1 \rangle + S_3 \langle w_3, w_3 \rangle ) \big].
\end{equation*}

\begin{enumerate}[label*=\alph*)]
\item Suppose $\mathbbm{1}_{(y_i^c <w_3+ y_i^f)} = 1$. Then
\begin{equation*}
\Psi_\alpha(x_i,y^c_i,y^f_i)  = \sup_{w_1 \in \mathbb{R}^n, w_3 \in B_n^2} \big[ \langle (w_1+x_i)^+,  y^c_i \rangle 
- \alpha ( \langle w_1, w_1 \rangle + S_3 \langle w_3, w_3 \rangle ) \big].
\end{equation*}
Recall $\langle (w_1 + x_i), y_i^c \rangle = (w_{1\tau_2} + x_{i\tau_2})$. 
Also recall $\tau_1$ and $\tau_2$ are associated with $y^f_i$.
Let $\tau_{2,c}$ denote the default time (index) for $y_i^c$.
The default time constraint implies $\tau_{2,c} < \tau_1$. Therefore $\tau_{2,c} > 0$ and $\|y_i^c\| = 1$.
Note the only positive component of $\|y^c_i\|$ is $\tau_{2,c}$. 
Hence set $w^*_{1\tau} = 0 \: \forall \tau \neq \tau_{2,c}$. Simplify further to get
\begin{equation*}
\Psi_\alpha(x_i,y^c_i,y^f_i)  = \sup_{w_{1\tau_{2,c}} \in \mathbb{R},w_3 \in B_n^2} \big[ (w_{1\tau_{2,c}}+x_{i\tau_{2,c}})^+ - \alpha ( ( w_{1\tau_{2,c}} )^2 + S_3 K^{3a} ) \big]
\end{equation*}
where $K^{3a} := ( \mathbbm{1}_{\{ \tau_1 \neq 0 \} } + \mathbbm{1}_{\{ \tau_2 \neq 0 \} } )$ = 1, following logic in \text{Case 2b)} above.
Evaluating at the critical point $w^{*}_{1\tau_{2,c}} = \frac{1}{2\alpha} \in \mathbb{R}$ gives
\begin{equation*}
\sup_{w_{1\tau_{2,c}} \in \mathbb{R}} \big[ (w_{1\tau_{2,c}}+ x_{i\tau_{2,c}})^+  - \alpha (w_{1\tau_{2,c}}^2 ) \big] = \big[\frac{1}{4\alpha} + x_{i\tau_{2,c}}\big]^+.
\end{equation*}
Therefore one can write
\begin{equation*}
\Psi_\alpha(x_i,y^c_i,y^f_i) =  \big[ \frac{1}{4\alpha} + x_{i\tau_{2,c}} \big]^+  - \alpha S_3 K^{3a}.
\end{equation*}
This can also be expressed as
\begin{equation*}
\Psi^{3a}_\alpha(x_i,y^c_i,y^f_i) = \bigg[ \big[\frac{1}{4 \alpha} + \langle x_i, y_i^c \rangle \big]^+ - \alpha S_3 K^{3a} \bigg].
\end{equation*}

\item Suppose $\mathbbm{1}_{(w_3+y_i^f <  y_i^c)} = 1$. Then
\begin{equation*}
\Psi_\alpha(x_i,y^c_i,y^f_i)  = \sup_{w_1 \in \mathbb{R}^n, w_3 \in B_n^2} \big[  \langle (w_1+x_i)^-, w_3+y^f_i \rangle 
- \alpha ( \langle w_1, w_1 \rangle + S_3 \langle w_3, w_3 \rangle ) \big].
\end{equation*}
Recall $\langle (w_1 + x_i), (w_3+y_i^f) \rangle \, = \, (w_{1\tau_1} + x_{i\tau_1})$. 
Also recall $\tau_1$ and $\tau_2$ are associated with $y^f_i$.
Let $\tau_{2,c}$ denote default time (index) for $y_i^c$.
The default time constraint implies $\tau_1 < \tau_{2,c}$.
Therefore $\tau_1 > 0$ and
\begin{equation*}
\Psi_\alpha(x_i,y^c_i,y^f_i) = \sup_{w_1 \in \mathbb{R}^n, 0 < \tau_1 < \tau_{2,c}, \tau_1 \neq \tau_2} \big[  (w_{1\tau_1} + x_{i\tau_1})^-  - \alpha ( (w_{1\tau_1})^2 + S_3 K^{3b} )  \big]
\end{equation*}
where $K^{3b} := ( 1 + \mathbbm{1}_{\{ \tau_2 \neq 0 \} } )$.
Following the calculations in $\text{Case} \: \text{2b})$ above, conclude that
\begin{equation*}
 \Psi_\alpha(x_i,y^c_i,y^f_i)  = \bigg[ \mathbbm{1}_{\{(x_{i\tau_1^*}<-\frac{1}{2\alpha}) \vee (x_{i\tau_1^*}>0)\}} \big[\frac{1}{4\alpha} + x_{i\tau_1^*}\big]^-  -
      \mathbbm{1}_{\{-\frac{1}{2\alpha} \leq x_{i\tau_1^*} \leq 0\}} \big[ \alpha (x_{i\tau_1^*})^2 \big] - \alpha S_3 K^{3b} \bigg].
\end{equation*}
Furthermore, $\tau_1^{*}$ is determined as
\begin{equation*}
\tau_1^{*} = \argmax_{0 < \tau_1 < \tau_2^c, \tau_1 \neq \tau_2} [ x_{i\tau_1} ].
\end{equation*}

Therefore one can write
\begin{equation*}
\Psi^{3b}_\alpha(x_i,y^c_i,y^f_i) =  \bigg[ \mathbbm{1}_{\{(x_{i\tau_1^*}<-\frac{1}{2\alpha}) \vee (x_{i\tau_1^*}>0)\}} \big[\frac{1}{4\alpha} + x_{i\tau_1^*}\big]^-  -
      \mathbbm{1}_{\{-\frac{1}{2\alpha} \leq x_{i\tau_1^*} \leq 0\}} \big[ \alpha (x_{i\tau_1^*})^2 \big] - \alpha S_3 K^{3b} \bigg].
\end{equation*}
This can also be expressed as
\begin{equation*}
\Psi^{3b}_\alpha(x_i,y^c_i,y^f_i) =  \bigg[ \mathbbm{1}_{\{(x_{i\tau_1^*}<-\frac{1}{2\alpha}) \vee (x_{i\tau_1^*}>0)\}} \big[\frac{1}{4\alpha} + \langle x_i, y_i^f \rangle + (x_{i\tau_1^*} - x_{i\tau_2}) \big]^-  -
      \mathbbm{1}_{\{-\frac{1}{2\alpha} \leq x_{i\tau_1^*} \leq 0\}} \big[ \alpha (x_{i\tau_1^*})^2 \big] - \alpha S_3 K^{3b} \bigg].
\end{equation*}

Finally, to sum up Case 3, considering parts a) and b), let us write:
\begin{equation*}
\Psi_\alpha^{3}(x_i,y^c_i,y^f_i)  = \mathbbm{1}_{(y_i^c < w_3+y_i^f)} \Psi_\alpha^{3a}(x_i,y^c_i,y^f_i) + \mathbbm{1}_{(w_3+y_i^f < y_i^c)} \Psi_\alpha^{3b}(x_i,y^c_i,y^f_i).
\end{equation*}
\end{enumerate}

$
\mathbf{Case \, 4}
$
\quad Suppose $w_2^{*} \neq 0, w_3^{*} \neq 0$.\\
Then $w_2^*$ has +1 in position $\tau_{1,c}^*$ and -1 in position $\tau_{2,c}$, where $\tau_{j,c} = 0$ means the value $\pm 1$ does not occur. \\
Furthermore, $\tau_{1,c}^* \neq \tau_{2,c}$ otherwise $w_2^* = 0$.\\
And $w_3^{*}$ has +1 in position $\tau_{1,f}^*$ and -1 in position $\tau_{2,f}$, where $\tau_{j,f} = 0$ means the value $\pm 1$ does not occur. \\
Furthermore, $\tau_{1,f}^* \neq \tau_{2,f}$ otherwise $w_3^* = 0$.\\
\begin{equation*}
\begin{split}
\Psi_\alpha(x_i,y^c_i,y^f_i)  = \sup_{w_1 \in \mathbb{R}^n, w_2 \in {B^2_n},w_3 \in {B^2_n}} \bigg[ \langle (w_1+x_i)^+, \mathbbm{1}_{\{w_2+y^c_i < w_3+y^f_i\}} w_2+y^c_i \rangle 
+  \langle (w_1+x_i)^-, \mathbbm{1}_{\{w_3+y^f_i < w_2+y^c_i\}} w_3+y^f_i \rangle \\
- \alpha ( \langle w_1, w_1 \rangle + S_3 \langle w_2, w_2 \rangle + S_3 \langle w_3, w_3 \rangle ) \bigg].
\end{split}
\end{equation*}

\begin{enumerate}[label*=\alph*)]
\item Suppose $\mathbbm{1}_{(w_2+y_i^c <w_3+ y_i^f)} = 1$. Then
\begin{equation*}
\Psi_\alpha(x_i,y^c_i,y^f_i)  = \sup_{w_1 \in \mathbb{R}^n, w_2 \in B_n^2,w_3 \in B_n^2} \big[ \langle (w_1+x_i)^+,  w_2 + y^c_i \rangle 
- \alpha ( \langle w_1, w_1 \rangle + S_3 \langle w_2, w_2 \rangle + S_3 \langle w_3, w_3 \rangle ) \big].
\end{equation*}
Recall $\langle (w_1 + x_i), (w_2+y_i^c) \rangle = (w_{1\tau_{1,c}} + x_{i\tau_{1,c}})$. 
The default time constraint implies $\tau_{1,c} < \tau_{1,f}$.
Therefore $\tau_{1,c} > 0$. \\
The structure of finite set $B^2_n$ implies
\begin{equation*}
\Psi_\alpha(x_i,y^c_i,y^f_i) = \sup_{w_1 \in \mathbb{R}^n, 0 < \tau_{1,c} < \tau_{1,f}, \tau_{1,c} \neq \tau_{2,c}} \big[  (w_{1\tau_{1,c}} + x_{i\tau_{1,c}})^+  - \alpha ( \langle w_1, w_1 \rangle + S_3 \langle w_2, w_2 \rangle + S_3 \langle w_3, w_3 \rangle ) \big].
\end{equation*}
Observe the only positive component for $w_1 \in \mathbb{R}^n$ in $\sup$ above is $\tau_{1,c}$.\\
\begin{equation*}
\sup_{w_1 \in \mathbb{R}^n} \big[ (w_{1\tau_{1,c}}+x_{i\tau_{1,c}})^+ - \alpha \langle w_1, w_1 \rangle \big] = \sup_{w_{1\tau_{1,c}} \in \mathbb{R}} \big[ (w_{1\tau_{1,c}}+x_{i\tau_{1,c}})^+ - \alpha (w_{1\tau_{1,c}}^2 ) \big].
\end{equation*}
Evaluating at the critical point $w^{*}_{1\tau_{1,c}} = \frac{1}{2\alpha} \in \mathbb{R}$ for the above quadratic gives
\begin{equation*}
\sup_{w_{1\tau_{1,c}} \in \mathbb{R}} \big[ (w_{1\tau_{1,c}}+ x_{i\tau_{1,c}})^+  - \alpha (w_{1\tau_{1,c}}^2 ) \big] = \big[\frac{1}{4\alpha} + x_{i\tau_{1,c}}\big]^+.
\end{equation*}
Therefore one can write
\begin{equation*}
\Psi_\alpha(x_i,y^c_i,y^f_i) = \max_{0 < \tau_{1,c} < \tau_{1,f}, \tau_{1,c} \neq \tau_{2,c}} \big[ \frac{1}{4\alpha} + x_{i\tau_{1,c}} \big]^+  - \alpha S_3 K^{4a}.
\end{equation*}
where $K^{4a} := ( \mathbbm{1}_{\{ \tau_{1,c} \neq 0 \} } + \mathbbm{1}_{\{ \tau_{2,c} \neq 0 \}} + \mathbbm{1}_{\{ \tau_{1,f} \neq 0 \} } + \mathbbm{1}_{\{ \tau_{2,f} \neq 0 \}} ) = ( 2 + \mathbbm{1}_{\{ \tau_{2,c} \neq 0 \} })$ following logic as in \text{Case 3a)} above.\\
Furthermore, $\tau_{1}^{*}$ is determined as
\begin{equation*}
\tau_{1}^{*} = \argmax_{0 < \tau_{1,c} < \tau_{1,f}, \tau_{1,c} \neq \tau_{2,c}} [ x^+_{i\tau_{1,c}} ].
\end{equation*}
Substituting back into expression for $\Psi_\alpha$ gives
\begin{equation*}
\Psi^{4a}_\alpha(x_i,y^c_i,y^f_i) = \bigg[ \big[ \frac{1}{4 \alpha} + x_{i\tau_1^{*}} \big]^+ - \alpha S_3 K^{4a} \bigg].
\end{equation*}
Let $\tau_2 = \tau_{2,c}$. Then this can also be expressed as
\begin{equation*}
\Psi^{4a}_\alpha(x_i,y^c_i,y^f_i) = \bigg[ \big[ \frac{1}{4 \alpha} + \langle x_i,y_i^c \rangle + ( x_{i\tau_1^*} - x_{i\tau_2} ) \big]^+ - \alpha S_3 K^{4a} \bigg].
\end{equation*}

\item Suppose $\mathbbm{1}_{(w_3+y_i^f <w_2+ y_i^c)} = 1$. Then
\begin{equation*}
\Psi_\alpha(x_i,y^c_i,y^f_i)  = \sup_{w_1 \in \mathbb{R}^n, w_2 \in B_n^2,w_3 \in B_n^2} \big[  \langle (w_1+x_i)^-, w_3+y^f_i \rangle 
- \alpha ( \langle w_1, w_1 \rangle + S_3 \langle w_2, w_2 \rangle + S_3 \langle w_3, w_3 \rangle ) \big].
\end{equation*}
Recall $\langle (w_1 + x_i), (w_3+y_i^f) \rangle = (w_{1\tau_{1,f}} + x_{i\tau_{1,f}})$. 
The default time constraint implies $\tau_{1,f} < \tau_{1,c}$.
Therefore $\tau_{1,f} > 0$.\\
The structure of finite set $B^2_n$ implies
\begin{equation*}
\Psi_\alpha(x_i,y^c_i,y^f_i) = \sup_{w_1 \in \mathbb{R}^n, 0 < \tau_{1,f} < \tau_{1,c}, \tau_{1,f} \neq \tau_{2,f}} \big[  (w_{1\tau_{1,f}} + x_{i\tau_{1,f}})^-  - \alpha ( \langle w_1, w_1 \rangle + S_3 \langle w_2, w_2 \rangle + S_3 \langle w_3, w_3 \rangle ) \big].
\end{equation*}

\begin{equation*}
\Psi_\alpha(x_i,y^c_i,y^f_i) = \sup_{w_1 \in \mathbb{R}^n, 0 < \tau_{1,f} < \tau_{1,c}, \tau_{1,f} \neq \tau_{2,f}} \big[  (w_{1\tau_{1,f}} + x_{i\tau_{1,f}})^-  - \alpha ( (w_{1\tau_{1,f}})^2 + S_3 K )  \big].
\end{equation*}
where $K := ( \mathbbm{1}_{\{ \tau_{1,c} \neq 0 \} } + \mathbbm{1}_{\{ \tau_{2,c} \neq 0 \} } + \mathbbm{1}_{\{ \tau_{1,f} \neq 0 \} } + \mathbbm{1}_{\{ \tau_{2,f} \neq 0 \} } ) = ( 2 + \mathbbm{1}_{\{ \tau_{2,f} \neq 0 \} })$ following logic as in \text{Case 4a)} above.\\
Following the calculations in $\text{Case} \: \text{3b})$ above, conclude that\\
\begin{equation*}
 \Psi_\alpha(x_i,y^c_i,y^f_i)  = \bigg[ \mathbbm{1}_{\{(x_{i\tau_1^*}<-\frac{1}{2\alpha}) \vee (x_{i\tau_1^*}>0)\}} \big[\frac{1}{4\alpha} + x_{i\tau_1^*}\big]^-  -
      \mathbbm{1}_{\{-\frac{1}{2\alpha} \leq x_{i\tau_1^*} \leq 0\}} \big[ \alpha (x_{i\tau_1^*})^2 \big] - \alpha S_3 K \bigg].
\end{equation*}
Furthermore, $\tau_1^{*}$ is determined as
\begin{equation*}
\tau_1^{*} = \argmax_{0 < \tau_{1,f} < \tau_{1,c}, \tau_{1,f} \neq \tau_{2,f}} [ x_{i\tau_{1,f}} ].
\end{equation*}

Therefore one can write
\begin{equation*}
\Psi^{4b}_\alpha(x_i,y^c_i,y^f_i) =  \bigg[ \mathbbm{1}_{\{(x_{i\tau_1^*}<-\frac{1}{2\alpha}) \vee (x_{i\tau_1^*}>0)\}} \big[\frac{1}{4\alpha} + x_{i\tau_1^*}\big]^-  -
      \mathbbm{1}_{\{-\frac{1}{2\alpha} \leq x_{i\tau_1^*} \leq 0\}} \big[ \alpha (x_{i\tau_1^*})^2 \big] - \alpha S_3 K^{4b} \bigg].
\end{equation*}
Let $\tau_2 = \tau_{2,f}$. Then this can also be expressed as
\begin{equation*}
\Psi^{4b}_\alpha(x_i,y^c_i,y^f_i) =  \bigg[ \mathbbm{1}_{\{(x_{i\tau_1^*}<-\frac{1}{2\alpha}) \vee (x_{i\tau_1^*}>0)\}} \big[\frac{1}{4\alpha} + \langle x_i, y_i^f \rangle + (x_{i\tau_1^*} - x_{i\tau_2}) \big]^-  -
      \mathbbm{1}_{\{-\frac{1}{2\alpha} \leq x_{i\tau_1^*} \leq 0\}} \big[ \alpha (x_{i\tau_1^*})^2 \big] - \alpha S_3 K^{4b} \bigg].
\end{equation*}

Finally, to sum up Case 4, considering parts a) and b), let us write:
\begin{equation*}
\Psi_\alpha^{4}(x_i,y^c_i,y^f_i)  = \mathbbm{1}_{(w_2+y_i^c < w_3+y_i^f)} \Psi_\alpha^{4a}(x_i,y^c_i,y^f_i) + \mathbbm{1}_{(w_3+y_i^f < w_2+y_i^c)} \Psi_\alpha^{4b}(x_i,y^c_i,y^f_i).
\end{equation*}

\end{enumerate}
\endgroup
\end{appendixproof}

\subsubsection{Outer Optimization Problem}
The goal now is to evaluate
\begin{equation*}
 \inf_{\alpha \geq 0} \:  F(\alpha) := \bigg[ \alpha \delta_3+ \frac{1}{N} \sum_{i=1}^{N} \Psi_{\alpha}(x_i,y^c_i,y^f_i) \bigg]  
\end{equation*}
where the $\Psi_\alpha$ functions are given as the solutions to Proposition 2.1.
Although the Lagrangian duality implies the convexity of $F(\alpha)$, due to its complexity, computational methods and solvers are used to evaluate this expression. Nonetheless, the solution can be expressed as below. Note that for $\delta_3 = 0$ one recovers the expression for original \text{CVA}\textsuperscript{B} given in Section 1.3.1.

\begin{theoremrep}
The primal problem \ref{eqn:primal3} has solution 
$\big[ \alpha^{*} \delta_3 + \frac{1}{N} \sum_{i=1}^{N} \Psi_{\alpha^{*}}(x_i,y^c_i,y^f_i) \big]$ \\
where $\alpha^{*} = \, \argmin_{\alpha \geq 0} \big[ \alpha \delta_3 + \frac{1}{N} \sum_{i=1}^{N} \Psi_{\alpha}(x_i,y^c_i,y^f_i) \big] $ and \,
$\Psi_{\alpha^{*}}(x_i,y^c_i,y^f_i) = \bigvee_{k=1}^4  \Psi_{\alpha^{*}}^{k}(x_i,y^c_i,y^f_i)$. \\ 
Expressed in terms of original BCVA, this says
\begin{equation*}
\sup_{\Phi \in \mathcal{U}_{\delta_3}(\Phi_N)} \mathbb{E}^\Phi [\langle X^+,Y^C \rangle + \langle X^-,Y^F \rangle ] = \mathbb{E}^{P_N} [\langle X^+,Y^C \rangle + \langle X^-,Y^F \rangle] + \alpha^{*} \delta_3 + \mathbb{E}^{P_N} \big[ \Psi_{\alpha^{*}}(X,Y^C,Y^F) - [\langle X^+,Y^C \rangle + \langle X^-,Y^F \rangle] \big]^+
\end{equation*}
where the additional terms represent a penalty due to uncertainty in probability distribution.
\end{theoremrep}
\begin{proofsketch}
This follows directly from the previous proposition. $\delta_3 = 0$ reduces to original BCVA. 
\end{proofsketch}
\begin{appendixproof}
This follows by direct substitution of $\alpha^{*}$ as characterized above into the dual problem \ref{eqn:dual3}.
\end{appendixproof}

\subsubsection{Recovering the Worst Case Distribution}
The process of recovering the worst case CVA distribution involves evaluating the $\argmin$ expressions given in Section 1.2.2. The procedure is a bit tedious but one can go through the various cases and subcases discussed in Proposition 2.1, and compute the value of the dual minimizer $\alpha^*$ as given in Theorem 2.1, to recover the worst case distribution $\{ (x^*_i, y^{c*}_i, y^{f*}_i) : i \in \{1,...,N+1\} \}$ for a given $\delta$. This procedure is done for a few concrete examples in Section 3.

\subsubsection{Discussion}
One limitation in the current approach is the omission of a risk neutral measure constraint on the underlying interest rate and credit default distributions that generate the portfolio exposure distributions described by the Wasserstein ball $\mathcal{U}_{\delta_3}(\Phi_N)$. It is not clear how to (either explicitly or implicitly) incorporate such a constraint in a solvable way. We highlight this as an opportunity for improvement and a direction for further research.
Empirical results for our worst case CVA studies are provided in Section 3. From the authors' perspective the computational study was illuminating to understand the magnitude and shape of worst case CVA profiles as a function of uncertainty.
Some recent work was done to map Wasserstein radii into lower and upper bounds on the distance between the true and empirical distributions. See the discussion on this topic in Section 3.2.
\endgroup

\begingroup
\setlength{\parindent}{0pt}
\subsection{FCA, FBA}
The robust FCA can be written as
\begin{equation*}\label{eqn:prima4}
\sup_{P \in \mathcal{U}_{\delta_1}(P_N)} \mathbb{E}^P [\langle Z^+, Y_{CF} \rangle]  \tag{P4}.
\end{equation*}
Similarly, the robust FBA can be written as 
\begin{equation*}\label{eqn:primal5}
 \sup_{Q \in \mathcal{U}_{\delta_2}(Q_N)} \mathbb{E}^Q [\langle Z^-, Y_{CF} \rangle]  \tag{P5}.
\end{equation*}
As such, the dual formulations and solutions to the above primal optimization problems are special cases of the solutions to the FVA optimization problems, to be described next.

\subsection{FVA}
\subsubsection{Inner Optimization Problem}
The robust FVA is
\begin{equation*}\label{eqn:primal6}
\sup_{\Phi \in \mathcal{U}_{\delta_3}(\Phi_N)} \mathbb{E}^\Phi [\langle Z, Y_{CF} \rangle]  \tag{P6}.
\end{equation*}

Similar to before, we use recent duality results, noting the inner product $\langle \: ; \rangle$ satisfies the upper semicontinuous condition of the Lagrangian duality theorem, and cost function $c_S$ satisfies the non-negative lower semicontinuous condition (see \citet{blanchetFirst} Assumptions 1 \& 2, \citet{Gao16}). Hence the dual problem (to sup above) can be written as
\begin{equation*}\label{eqn:dual6}
 \inf_{\alpha \geq 0} \: F(\alpha) := \bigg[ \alpha \delta_3 + \frac{1}{N} \sum_{i=1}^{N} \Psi_\alpha(z_i,y^{cf}_i) \bigg]  \tag{D6}
\end{equation*}
where 
\begin{equation*}
\Psi_\alpha(z_i,y^{cf}_i)  = \sup_{u \in \mathbb{R}^n, v \in {B^1_n}} [  \langle u,v \rangle - \alpha c_{S_3}((u,v),(z_i,y^{cf}_i)) ] = \sup_{u \in \mathbb{R}^n, v \in {B^1_n}} [  \langle u, v \rangle - \alpha( \langle u-z_i, u-z_i \rangle + S_3 \langle v-y^{cf}_i, v-y^{cf}_i \rangle ) ].
\end{equation*}


Now apply change of variables $w_1 = (u-z_i)$ and $w_2 = (v-y^{cf}_i)$ to get
\begin{equation*}
\Psi_\alpha(z_i,y^{cf}_i)  = \sup_{w_1 \in \mathbb{R}^n, w_2 \in {B^2_n}} [ \langle w_1+z_i, w_2+y^{cf}_i \rangle - \alpha ( \langle w_1, w_1 \rangle + S_3 \langle w_2, w_2 \rangle ) ]
\end{equation*}
where the sets $B^1_n$ and $B^2_n$ are defined as before. It turns out that $\Psi_\alpha$ can be expressed as original FVA plus the pointwise max of $(n+1)$ convex functions. The degenerate case $l=0$ is just a line of negative slope. The other $n$ cases are hyperbolas plus lines of negative slope. $\Psi_\alpha$ quantifies the adversarial move in FVA across both time and spatial dimensions while accounting for the cost via the $K$ terms. 


\begin{proprep}
We have \, $\Psi_\alpha(z_i,y^{cf}_i) = \, \langle z_i,y^{cf}_i \rangle + \big[  \frac{l^*}{4 \alpha} + \big(\sum_{k=1}^{l^*}z_{ik} - \sum_{k=1}^{\|y^{cf}_i\|_1} z_{ik}\big) - \alpha S_3 K \big]$ \\
where $l^{*} = \argmax_{l \geq 0} [ \frac{l}{4 \alpha} + \sum_{k=1}^{l} z_{ik} - \alpha S_3 K]$ and $l = \|w_2 + y^{cf}_i \|_1 \geq 0, \:  l \in \mathbb{Z^+}$.
Also $\|y^{cf}_i\|_1 \in \mathbb{Z^+}$, and  $K = | l - \| y^{cf}_i \|_1 | = \| w_2 \|_1 \geq 0, K \in \mathbb{Z^+}$.
Once $l^*$ is selected, $K := | l^* - \| y^{cf}_i \|_1 | = \| w_2^* \|_1$.
Alternatively, $\Psi_\alpha(z_i,y^{cf}_i) = \, \langle z_i,y^{cf}_i \rangle + \bigvee_{l=0}^n  h_\alpha(l)$ for \\ $h_\alpha(l) := \big[  \frac{l}{4 \alpha} + \big(\sum_{k=1}^{l}z_{ik} - \sum_{k=1}^{\|y^{cf}_i\|_1} z_{ik}\big) - \alpha S_3 K \big]$.
\end{proprep}
\begin{proofsketch}
This result follows from jointly maximizing the adversarial funding exposure $w_1$ and the survival time index $w_2$.  The structure of $B^2_n$ allows us to decouple this joint maximization and find the critical point to maximize the quadratic in $w_1$ and write down the condition to select the optimal survival time index $l^*$. Finally, consider the two cases $w_2 = 0$ and $w_2 \neq 0$ and take the max to arrive at the solution. The $K$ terms represent the cost associated with the worst case.
\end{proofsketch}
\begin{appendixproof}
\begingroup
\setlength{\parindent}{0pt}

The particular structure of $B^1_n$ and $B^2_n$ will be exploited to evaluate the $\sup$ above.
The analysis proceeds by considering different cases for optimal values $(w_1^{*}, w_2^{*})$.\\ \\
\medskip
$
\mathbf{Case \, 1}
$
\quad Suppose $w_2^{*} = 0 \implies  l = \| y^{cf}_i \|_1$. Then\\ 
\begin{equation*}
\Psi_\alpha(z_i,y^{cf}_i)  = \, \langle z_i,y^{cf}_i \rangle + \sup_{w_1 \in \mathbb{R}^n} [  \langle w_1, y^{cf}_i \rangle - \alpha \langle w_1, w_1 \rangle ].
\end{equation*}
Applying the Cauchy-Schwarz Inequality gives
\begin{equation*}
\Psi_\alpha(z_i,y^{cf}_i)  = \, \langle z_i,y^{cf}_i \rangle + \sup_{\| w_1 \|} [ \| w_1\| \| y^{cf}_i \| - \alpha \| w_1 \|^2 ].
\end{equation*}
Evaluating the critical point $\|w_1^{*}\| = \frac{\|y^{cf}_i\|}{2\alpha} \in \mathbb{R}_{+}$ for the quadratic gives
\begin{equation*}
\Psi_\alpha(z_i,y^{cf}_i)  = \, \langle z_i,y^{cf}_i \rangle + \frac{ \| y^{cf}_i \|^2}{4\alpha} = \langle z_i,y^{cf}_i \rangle + \frac{ \| y^{cf}_i \|_1}{4\alpha}.
\end{equation*} \\

$
\mathbf{Case \, 2}
$
\quad Now consider $w_2^{*} \neq 0 \implies l \neq \| y^{cf}_i \|_1 $. \\
Observe for $l = \|w_2 + y^{cf}_i \|_1 \geq 0$, \\
\begin{equation*}
\langle w_1+z_i, w_2 + y_i^{cf} \rangle = \sum_{k=1}^{l} (w_{1k} + z_{ik}).
\end{equation*}
The structure of finite set $B^2_n$ implies
\begin{equation*}
\Psi_\alpha(z_i,y^{cf}_i) = \sup_{w_1 \in \mathbb{R}^n, l \in \{0,\ldots,n\}, l \neq  \| y^{cf}_i \|_1} [ \sum_{k=1}^{l} (w_{1k} + z_{ik})  - \alpha ( \langle w_1, w_1 \rangle + S_3 K ) ].
\end{equation*}

Again, using that $B^2_n$ is a finite set, one can write
\begin{equation*}
\Psi_\alpha(z_i,y^{cf}_i) = \max_{ l \in \{0,\ldots,n\}, l \neq  \| y^{cf}_i \|_1} \sup_{w_1 \in \mathbb{R}^n} [ \sum_{k=1}^{l} (w_{1k} + z_{ik}) - \alpha ( \langle w_1, w_1 \rangle + S_3 K ) ].
\end{equation*}
Observing that only the first $l$ components of $w_1 \text{ inside the sup are positive gives } \forall k \in \{1,\ldots,l\}$
\begin{equation*}
\sup_{w_1 \in \mathbb{R}^n} [ \sum_{k=1}^{l} (w_{1k}) - \alpha \langle w_1, w_1 \rangle  ]  = l \times \sup_{w_{1k} \in \mathbb{R}} [ w_{1k} - \alpha ( w_{1k} )^2 ].
\end{equation*}
Evaluating at the critical point $w^{*}_{1k} = \frac{1}{2\alpha} \in \mathbb{R}_+$ for the above quadratic gives
\begin{equation*}
\sup_{w_{1k} \in \mathbb{R}} [ w_{1k} - \alpha (w_{1k}^2 ) ] = \frac{1}{4\alpha}.
\end{equation*}
Therefore one can write
\begin{equation*}
\Psi_\alpha(z_i,y^{cf}_i) = \max_{l \in \{0,\ldots,n\}, l \neq  \| y^{cf}_i \|_1} [ \frac{l}{4\alpha} + \sum_{k=1}^{l} (z_{ik}) - \alpha S_3 K ].
\end{equation*}

Furthermore, $l^{*}$ is determined as
\begin{equation*}
l^{*} = \argmax_{l \in \{0,\dots,n\}, l \neq  \| y^{cf}_i \|_1} [ \frac{l}{4\alpha} + \sum_{k=1}^{l} (z_{ik}) - \alpha S_3 K ].
\end{equation*}

Substituting back into expression for $\Psi_\alpha$ gives
\begin{equation*}
\Psi_\alpha(z_i,y^{cf}_i) = \, \langle z_i,y^{cf}_i \rangle + \bigg[  \frac{l^*}{4 \alpha} +  \bigg( \sum_{k=1}^{l^*} z_{ik}  - \sum_{k=1}^{\|y^{cf}_i\|_1} z_{ik} \bigg)  - \alpha S_3 K \bigg].
\end{equation*}

Finally, taking the max values for $\Psi_\alpha$ over cases $w_2^{*} = 0$ and $w_2^{*} \neq 0$ gives
\begin{equation*}
\Psi_\alpha(z_i,y^{cf}_i) = \, \langle z_i,y^{cf}_i \rangle + \bigg[\frac{ \| y^{cf}_i \|_1}{4 \alpha}\bigg] \vee  \bigg[  \frac{l^*}{4 \alpha} + \bigg( \sum_{k=1}^{l^*} z_{ik}  - \sum_{k=1}^{\|y^{cf}_i\|_1} z_{ik} \bigg) - \alpha S_3 K \bigg].
\end{equation*}
Observe that for $l^* = \| y^{cf}_i \|_1$, the last term in brackets $[ \, ; ]$ above evaluates to $\big[\frac{ \| y^{cf}_i \|_1}{4 \alpha}\big]$.
Let $l^*$ be determined as
\begin{equation*}
l^{*} = \argmax_{l \in \{0,\dots,n\}} [ \frac{l}{4\alpha} + \sum_{k=1}^{l} (z_{ik}) - \alpha S_3 K ]
\end{equation*}
and write
\begin{equation*}
\Psi_\alpha(z_i,y^{cf}_i) = \, \langle z_i,y^{cf}_i \rangle + \bigg[  \frac{l^*}{4 \alpha} + \bigg(\sum_{k=1}^{l^*}z_{ik} - \sum_{k=1}^{\|y^{cf}_i\|_1} z_{ik}\bigg) - \alpha S_3 K \bigg].
\end{equation*}

Alternatively, one can write
\begin{equation*}
\Psi_\alpha(z_i,y^{cf}_i) = \, \langle z_i,y^{cf}_i \rangle + \bigvee_{l=0}^n  \bigg[  \frac{l}{4 \alpha} + \bigg(\sum_{k=1}^{l}z_{ik} - \sum_{k=1}^{\|y^{cf}_i\|_1} z_{ik}\bigg) - \alpha S_3 K \bigg].
\end{equation*}

\endgroup
\end{appendixproof}

%
\subsubsection{Outer Optimization Problem}

The goal now is to evaluate
\begin{equation*}
 \inf_{\alpha \geq 0} \:  F(\alpha) := \bigg[ \alpha \delta_3 + \frac{1}{N} \sum_{i=1}^{N} \Psi_\alpha(z_i,y^{cf}_i) \bigg]  
\end{equation*}
where 
\begin{equation*}
\Psi_\alpha(z_i,y^{cf}_i)  = \, \langle z_i,y^{cf}_i \rangle + \bigvee_{l=0}^n h_\alpha(l) \:\: \text{for} \:\:  h_\alpha(l) := \big[  \frac{l}{4 \alpha} + \big(\sum_{k=1}^{l}z_{ik} - \sum_{k=1}^{\|y^{cf}_i\|_1} z_{ik}\big) - \alpha S_3 K \big].
\end{equation*}
The convexity of the objective function $F(\alpha)$ simplifies the task of solving this optimization problem. The first order optimality condition suffices. As $\Psi_\alpha$ and hence $F(\alpha)$ may have non-differentiable kinks due to the max functions, $\vee$, we characterize the optimality condition via subgradients. In particular, we look for $\alpha^{*} \geq 0$ such that $0 \in \partial F(\alpha^{*})$. Inspection of the asymptotic properties of $\Psi_\alpha$ and its subgradients reveals that $\partial F(\alpha)$ will cross zero (as $\alpha$ sweeps from $0$ to $\infty$) and hence $\alpha^{*} \geq 0$. Note that for $\delta_3 = 0$ one recovers the expression for original \text{FVA} given in Section 1.3.3.


\begin{proprep}
Let $\alpha^{*} \in \,  \left\{ \alpha \geq 0: 0 \in \partial F(\alpha) \right\} \\ where \,\, \partial \Psi_\alpha = \mathbf{Conv} \cup \left\{ \partial h_\alpha(l) \: | \: \langle z_i,y^{cf}_i \rangle + h_\alpha(l) = \Psi_\alpha ; \, l \in \{0,\dots,n\} \right\}$
and $\partial F(\alpha) = \delta_3 + \frac{1}{N} \sum_{i=1}^N \partial \Psi_\alpha$\,.\\
\end{proprep}

\begin{proofsketch}
This follows from application of standard properties of subgradients as well as inspection of the asymptotic properties of $\Psi_\alpha$ and $\partial \Psi_\alpha$. For $\alpha$ sufficiently small, $\Psi_\alpha$ has a large positive value and $\partial \Psi_\alpha$ has a large negative derivative. For $\alpha$ sufficiently large, for optimal $l^*$, either $l^* = 0 \implies 0 \in \partial \Psi_\alpha$ or $l^* = \| y^{cf}_i \|_1 > 0 \implies \partial \Psi_\alpha$ approaches zero $\implies \partial F(\alpha)$ crosses zero.
\end{proofsketch}

\begin{appendixproof}
\begingroup
\setlength{\parindent}{0pt}

This follows from standard application of properties of convex functions and subgradients. 
First note that function $h_\alpha$ is convex in $\alpha$ since (for fixed $l$) it is the sum of a hyperbola plus a constant plus a negative linear term. 
So then $\Psi_\alpha$ is convex since it is the pointwise max of a finite set of convex functions plus a constant. 
Using properties of subgradients, one can write $\partial \Psi_\alpha = \mathbf{Conv} \cup \{ \partial h_\alpha(l) \: | \: \langle z_i,y^{cf}_i \rangle + h_\alpha(l) = \Psi_\alpha ; \, l \in \{0,\dots,n\} \}$. 
Furthermore $F(\alpha)$ is convex in $\alpha$ since it is a linear term plus a sum of convex functions, so one can write $\alpha^{*} \in \,  \left\{ \alpha : 0 \in \partial F(\alpha) \right\}$ and it follows that  $\partial F(\alpha) = \delta_3 + \frac{1}{N} \sum_{i=1}^N \partial \Psi_\alpha$.
Finally, we argue that $\alpha^{*} \geq 0$. 
For $\alpha > 0$ sufficiently small, $\exists \, z  < -\delta_3$ such that $z \in \partial \Psi_\alpha$ and for $\alpha > 0$ sufficiently large, $\exists \, z > -\delta_3$ such that $z \in \partial \Psi_\alpha$. 
To elaborate, for $\alpha > 0$ sufficiently large, $\| y^{cf}_i \|_1 > 0 \implies l^* = \| y^{cf} \|_1 \implies K = 0 \implies \exists \, z > -\delta_3$ such that $z \in \partial \Psi_\alpha$.
To elaborate, for $\alpha > 0$ sufficiently large, $\| y^{cf}_i \|_1 = 0 \implies l^* = 0 \implies K = 0, \Psi_\alpha = 0, 0 = z > -\delta_3$ such that $z \in \partial \Psi_\alpha$.
Hence we deduce $\partial F(\alpha)$ crosses the origin ( as $\alpha$ sweeps from $0$ to $\infty$ ). 
\endgroup
\end{appendixproof}


\begin{theoremrep}
The primal problem \ref{eqn:primal6} has solution 
$\big[ \alpha^{*} \delta_3 + \frac{1}{N} \sum_{i=1}^{N} \Psi_{\alpha^{*}}(z_i,y^{cf}_i) \big]$ \\
where $\alpha^{*} \in \,  \left\{ \alpha \geq 0: 0 \in \partial F(\alpha) \right\}$ and
$\Psi_{\alpha^{*}}(z_i,y^{cf}_i) = \, \langle z_i,y^{cf}_i \rangle + \bigvee_{l=0}^n h_{\alpha^{*}}(l) \:\: for \:\:  h_{\alpha^{*}}(l) := \big[  \frac{l}{4 \alpha^{*}} + \big(\sum_{k=1}^{l}z_{ik} - \sum_{k=1}^{\|y^{cf}_i\|_1} z_{ik}\big) - \alpha^{*} S_3 K \big]$. 
Expressed in terms of original FVA, this says
\begin{equation*}
\sup_{\Phi \in \mathcal{U}_{\delta_3}(\Phi_N)} \mathbb{E}^\Phi [\langle Z,Y_{CF} \rangle] = \mathbb{E}^{\Phi_N} [\langle Z,Y_{CF} \rangle] + \alpha^{*} \delta_3 +   \mathbb{E}^{\Phi_N} \big[ \bigvee_{l=0}^n \frac{l}{4 \alpha^{*}} + \big(\sum_{k=1}^{l}Z_k - \sum_{k=1}^{\|Y_{CF}\|_1} Z_k \big) - \alpha^{*} S_3 K  \big]
\end{equation*}
where the additional terms represent a penalty due to uncertainty in probability distribution.
\end{theoremrep}
\begin{proofsketch}
This follows directly from the previous two propositions. $\delta_3 = 0$ reduces to original FVA. 
\end{proofsketch}
\begin{appendixproof}
This follows by direct substitution of $\alpha^{*}$ as characterized in Proposition 2.3 into Proposition 2.2 and then the dual problem \ref{eqn:dual6}.
\end{appendixproof}

\subsubsection{Recovering the Worst Case Distribution}
The process of recovering the worst case FVA distribution is similar to that for CVA. In fact, for the FVA case, the procedure is a bit simpler since there are less cases and subcases to consider to recover $\{ x^*_i, y^{cf*}_i : i \in \{1,...,N+1\} \}$. The steps to recover the dual minimizer $\alpha^*$ are the same. This procedure is done for a few concrete examples in Section 3.

\subsubsection{Discussion}
The comments regarding incorporation of risk neutral measure constraint for the robust CVA problem formulations apply for the robust FVA problem formulations as  well. Empirical results for the worst case FVA studies are provided in Section 3. Similar to CVA, from the authors' perspective the computational study was illuminating to understand the magnitude and shape of worst case FVA profiles as a function of uncertainty.

\endgroup

\section{Computational Study: Robust XVA and Wrong Way Risk}
This computational study uses the Matlab Financial Instruments Toolbox and extends WWR portfolio analysis \cite[section 5.3]{brigo2013counterparty} to consider uncertainty in probability distribution. Other key concepts that will be discussed in this section include suitable choice for Wasserstein radius $\delta$, calibration of scale factor $S_3$, and choice of units for exposures. The studies in this section will investigate (and quantify) worst case bilateral CVA and FVA for different market environments and portfolios of interest rate swaps. For CVA, the current swaps market data (see below) will be used in conjunction with Monte Carlo simulation of a market calibrated one factor Hull-White model for interest rates. The counterparty credit curve selection will vary between investment grade and high yield. For FVA, the funding spreads and volatility data is taken from Markit. The swaps portfolios are shown as well. All calculations are done in Matlab using the financial instruments toolbox \citep{Matlab19}.

\subsection{Market Data}
As of April 20, 2020, the 5y par interest rate swap rate is 0.47\% (on Bloomberg). The full interest rate swaps curve is shown in Table 2. All market data displayed below is for this date.
\begin{table}[h]
\begin{center}
\caption{Swap Rates}
\begin{tabular}{ |c|c|c|c|c|c|c|c| }
 \hline
Swap Tenor & 1y & 2y & 3y & 5y & 7y & 10y & 30y \\
 \hline
Swap Rate & 0.515\% & 0.409\% & 0.401\% & 0.470\% & 0.569\% & 0.691\% & 0.855\% \\
 \hline
\end{tabular}
\end{center}
\end{table}

\noindent Bloomberg shows the interest rate swaption volatility matrix (with option expirations as rows and swap tenors as columns).
\begin{table}[h]
\begin{center}
\caption{Swaption Normal Volatilities}
\begin{tabular}{ |c|c|c|c|c|c| }
 \hline
Exp / Tenor & 2y & 3y & 5y & 7y & 10y \\
 \hline
2y & 0.520\% & 0.542\% & 0.601\% & 0.631\% & 0.680\% \\
\hline
3y & 0.577\% & 0.592\% & 0.622\% & 0.640\% & 0.671\% \\
\hline
5y & 0.637\% & 0.637\% & 0.637\% & 0.643\% & 0.652\% \\
\hline
7y & 0.640\% & 0.639\% & 0.636\% & 0.636\% & 0.636\% \\
\hline
10y & 0.639\% & 0.633\% & 0.624\% & 0.618\% & 0.612\% \\
 \hline
\end{tabular}
\end{center}
\end{table}


\noindent Furthermore, Markit shows U.S. CDX investment grade and high yield 5y credit default swap spreads as in Table 4. The firm and counterparty investment grade credit spreads are set to 100 and 150 basis points respectively. The high yield credit spreads are shown in Table 5. Referencing Markit funding spreads, the funding spread curves are shown in Table 6. Unavailable quotes for high yield spreads are displayed as ``N/A". This term structure of funding spreads is used for the FVA analysis. Funding spread lognormal volatility is set to exponential decay. For investment grade it decays from 85\% down to about 31\% in 10 years. For high yield it decays from 35\% down to about 13\% in 10 years.

\begin{table}[h]
\begin{center}
\caption{5y CDS Spreads}
\begin{tabular}{ |c|c|c| }
 \hline
CDX Index & IG & HY \\
 \hline
CDS Spread & 0.933\% & 6.432\% \\
 \hline
\end{tabular}
\end{center}
\end{table}

\begin{table}[H]
\begin{center}
\caption{High Yield Counterparty Credit Spreads}
\begin{tabular}{ |c|c|c|c|c|c|c|c|c|c|c| }
 \hline
CDS Tenor & 1y & 2y & 3y & 4y & 5y & 6y & 7y & 8y & 9y & 10y \\
 \hline
HY Spread & 6.00\% & 5.75\% & 5.50\% & 5.25\% & 5.00\% & 4.75\% & 4.50\% & 4.25\% & 4.00\% & 3.75\% \\
 \hline
\end{tabular}
\end{center}
\end{table}

\begin{table}[H]
\begin{center}
\caption{Funding Spreads}
\begin{tabular}{ |c|c|c|c|c|c|c| }
 \hline
Funding Tenor & 1y & 2y & 3y & 5y & 7y & 10y \\
 \hline
IG Spread & 0.54\% & 0.81\% & 0.81\% & 0.88\% & 1.01\% & 1.14\% \\
 \hline
HY Spread & N/A & N/A & 8.02\% & 6.72\% & 7.08\% & 6.36\% \\
 \hline
\end{tabular}
\end{center}
\end{table}

\noindent The swaps portfolios for the CVA and FVA studies are shown in Tables 7 and 8. All 10 swaps are used for the 30y Monte Carlo simulation for CVA. The last 4 are capped at 10y maturity for FVA, as we have (only) 10y of funding market data.
\begin{table}[h]
\begin{center}
\caption{CVA Swaps Portfolio}
\begin{tabular}{ |c|c|c|c|c|c| }
 \hline
Issued & Notional & Maturity & Rec / Pay Fixed & Coupon & Freq \\
 \hline
4/20/20 & 10 & 4/20/21 & Rec & 0.51\% & quarterly \\
\hline
4/20/20 & 10 & 4/20/22 & Pay & 0.41\% & quarterly \\
\hline
4/20/20 & 10 & 4/20/23 & Pay & 0.40\% & quarterly \\
\hline
4/20/20 & 10 & 4/20/25 & Rec & 0.47\% & quarterly \\
\hline
4/20/20 & 10 & 4/20/27 & Pay & 0.57\% & quarterly \\
 \hline
4/20/20 & 10 & 4/20/30 & Rec & 0.69\% & quarterly \\
\hline
4/20/20 & 10 & 4/20/35 & Rec & 0.74\% & quarterly \\
\hline
4/20/20 & 10 & 4/20/40 & Rec & 0.83\% & quarterly \\
\hline
4/20/20 & 10 & 4/20/45 & Pay & 0.83\% & quarterly \\
\hline
4/20/20 & 10 & 4/20/50 & Pay & 0.85\% & quarterly \\
 \hline
\end{tabular}
\end{center}
\end{table}

\begin{table}[h]
\begin{center}
\caption{FVA Swaps Portfolio}
\begin{tabular}{ |c|c|c|c|c|c| }
 \hline
Issued & Notional & Maturity & Rec / Pay Fixed & Coupon & Freq \\
 \hline
4/20/20 & 100 & 4/20/21 & Pay & 0.51\% & quarterly \\
\hline
4/20/20 & 100 & 4/20/22 & Rec & 0.41\% & quarterly \\
\hline
4/20/20 & 100 & 4/20/23 & Rec & 0.40\% & quarterly \\
\hline
4/20/20 & 100 & 4/20/25 & Pay & 0.47\% & quarterly \\
\hline
4/20/20 & 100 & 4/20/27 & Rec & 0.57\% & quarterly \\
 \hline
4/20/20 & 100 & 4/20/30 & Pay & 0.69\% & quarterly \\
\hline
4/20/20 & 100 & 4/20/30 & Pay & 0.74\% & quarterly \\
\hline
4/20/20 & 100 & 4/20/30 & Pay & 0.83\% & quarterly \\
\hline
4/20/20 & 100 & 4/20/30 & Rec & 0.83\% & quarterly \\
\hline
4/20/20 & 100 & 4/20/30 & Rec & 0.85\% & quarterly \\
 \hline
\end{tabular}
\end{center}
\end{table}

\subsection{Suitable Choice for Wasserstein Radius}

A natural question to ask when computing worst case XVA is how to interpret the size of the Wasserstein radius $\delta$. A discussion of some key results is given in \cite[Section 3]{Carlsson2018}. For this study, we adopt a fairly straightforward approach to compute upper and lower bounds for the expected Wasserstein distance between the empirical and true distributions. A rough procedure for selecting $\delta$ involves sampling two independent data sets $D_1 \text{\,and\,} D_2$, and setting $\delta = \alpha c^*$ where $\alpha \in [1/2,1]$ and $c^*$ denotes the cost of the minimum bipartite matching between $D_1 \text{\,and\,} D_2$ \citep{Carlsson2018}, \citep{radius12}. This approach relies on the following theorem referenced in \citet{Carlsson2018} and established in \citet{radius12}.
\begin{theorem*}
Let $\hat{f}_1$ and $\hat{f}_2$ denote empirical distributions associated with two sets of independent samples of $n$ points from a distribution $f$. Then
\[
\frac{1}{2} \mathbb{E} \; [D_c(\hat{f}_1,\hat{f}_2)] \leq \mathbb{E} \; [D_c(f,\hat{f}_1)] \leq \mathbb{E} \; [D_c(\hat{f}_1,\hat{f}_2)].
\]
\end{theorem*}
As such, our approach is to sample two indepedent data sets $D_1$ and $D_2$ of portfolio exposures and default times and compute lower and upper bounds $\delta^l := \frac{1}{2} \mathbb{E} \; [D_c(\hat{f}_1,\hat{f}_2)]$ and $\delta^u := \mathbb{E} \; [D_c(\hat{f}_1,\hat{f}_2)]$ for the expected Wasserstein distance between the empirical and true distributions. Given these bounds, one can compute the corresponding lower and upper bounds on the worst case XVA risk metrics and exposure and default time distributions. \par
Constructing the bounds $\delta^l$ and $\delta^u$ in this way builds in a dependency on the units of portfolio exposures (e.g. millions of dollars) and units in the time dimension (e.g. years), through the computation of $D_c(\hat{f}_1,\hat{f}_2)$ and the calibration of the scale factor $S_3$ (see Section 3.3 below for this). Such a dependency is desirable to assign ``units" to $\delta$ as well as to conduct relative value analysis across portfolios. See Section 3.4 below for more commentary on choice of units for exposures.

\subsection{Calibration of Scale Factors}
\subsubsection{Calibration of $S_3$ for CVA}
The scale factor $S_3$ represents a scaling for changes to default times. A suitable choice for $S_3$ is one that charges an appropriate cost for this. Let us think about what a change in default time means in the context of CVA. For a fixed path with index $i$, and exposure vector $x^\pm_i$, changing the default time from $\tau_2$ to $\tau_1$ changes the value of the realized exposure from $x^\pm_{i\tau_2}$ to $x^\pm_{i\tau_1}$ upon default. 
A reasonable value for $S_3$, call it $s_3$, for this particular path, might be $\|{x}^\pm_{i\tau_1} - {x}^\pm_{i\tau_2}\|_\infty$ where $\tau_1,\tau_2 \in \{1,...,n\}$.
Now let us generalize this to average over all paths $i \in \{1,...,N\}$ in our empirical distribution $\Phi_N$. Let $\overline{{x}^\pm_{\tau}}$ denote $\frac{1}{N} \sum_ {i=1}^N x^\pm_{i\tau}$, the average exposure at default time $\tau$. Substituting average exposures into our previous expression gives the relation $S_3 := \| \overline{{x}^\pm_{\tau_1}} - \overline{{x}^\pm_{\tau_2}} \|_\infty$. Let us use this as our working definition for $S_3$ for unilateral CVA, DVA. Calibration is straightforward given $\Phi_N$, the set of sample paths $\{ (x_i, y^c_i, y^f_i ) : i \in \{ 1,...,N \} \}$. For bilateral CVA, take the average over the unilateral CVA and DVA scale factors, namely $S_3 := \frac{1}{2} ( \| \overline{{x}^+_{\tau_1}} - \overline{{x}^+_{\tau_2}} \|_\infty + \| \overline{{x}^-_{\tau_1}} - \overline{{x}^-_{\tau_2}} \|_\infty)$.


\subsubsection{Calibration of $S_3$ for FVA}
Let us the follow the approach above for FVA. For a fixed path with index $i$, the funding exposure vector is $z^\pm_i$ and the incremental change is $\Delta z^\pm_{i\tau_2}$. A reasonable value for $S_3$, call it $s_3$, for this particular path, might be $\|{z}^\pm_{i\tau_1} - {z}^\pm_{i\tau_2}\|_\infty$. Substituting average exposures into this expression gives the relation $S_3 := \| \overline{{z}^\pm_{\tau_1}} - \overline{{z}^\pm_{\tau_2}} \|_\infty$. Let us use this as our working definition for $S_3$ for FCA, FBA. Calibration is straightforward given $\Phi_N$, the set of sample paths $\{ (z_i, y^{cf}_i) : i \in \{ 1,...,N \} \}$. For FVA, take the average over the FCA and FBA scale factors, namely $S_3 := \frac{1}{2} ( \| \overline{{z}^+_{\tau_1}} - \overline{{z}^+_{\tau_2}} \|_\infty + \| \overline{{z}^-_{\tau_1}} - \overline{{z}^-_{\tau_2}} \|_\infty)$.

\subsection{Choice of Units for Exposures}
Standardizing the units across portfolios is useful for relative value analysis. The choice of units for exposures (e.g. millions of dollars) and default times (e.g. decimal years) is up to the user, although we recommend these conventions, and use them in our analysis in this section. Note that different choices of units will lead to calibrated different values for $S_3$ for BCVA and FVA. There is no \textit{one} choice for units (as in regression analysis, for example) although consistency is recommended as a good practice. The same comments apply for the choices of time frequency and time horizon for the robust XVA analysis.

\subsection{Definitions for Exposure Calculations}
The definitions for the various exposure calculations plotted in Section 3.6 for CVA and DVA are given in Table 9 below. For FVA calculations (FCA and FBA), plotted in Section 3.7, replace portfolio exposures $V^+$ and $V^-$ with funding exposures $Z^+$ and $Z^-$ respectively.
\renewcommand{\arraystretch}{1.5}
\begin{table}[H]
\normalsize
\begin{center}
\caption{CVA Exposure Calculations}
\begin{tabular}{ |c|l|l| }
 \hline
\textit{Term} & \text{CVA}\textsuperscript{U} & \text{DVA}\textsuperscript{U} \\
 \hline
$\text{EE}(t)$ & $\mathbb{E} [ V^+(t) ]$ & $\mathbb{E} [ V^-(t) ]$ \\ 
\hline
$\text{PFE}_\alpha(t)$ & $\inf \; \{ x \in \mathbb{R} : \alpha \leq F_{V^+(t)}(x) \}$ & $\inf \; \{ x \in \mathbb{R} : \alpha \leq F_{V^-(t)}(x) \}$ \\ 
 \hline
$\text{EPE}(t)$ & $\frac{1}{T} \int_0^T \text{EE}(t) dt$ & $\frac{1}{T} \int_0^T \text{EE}(t) dt$ \\ 
\hline
$\text{EffEE}(t)$ & $\max \; \{ \text{EE}(\tau) : \tau \in [0,t] \}$ & $\max \; \{ \text{EE}(\tau) : \tau \in [0,t] \}$ \\ 
\hline
$\text{EffEPE}(t)$ & $\frac{1}{T} \int_0^T \text{EffEE}(t) dt$ & $\frac{1}{T} \int_0^T \text{EffEE}(t) dt$ \\ 
\hline
\end{tabular} 
\end{center} 
\end{table}
\renewcommand{\arraystretch}{1}

\subsection{Bilateral CVA}
\subsubsection{Investment Grade Counterparty and Firm}
The swaps portfolio shown in Table 7 is used for this analysis. The portfolio consists of ten par coupon interest rate swaps, with a mix of receving fixed and paying fixed swaps at different maturities. The investment grade firm and counterparty credit spreads are set to 100 and 150 basis points respectively. The calibrated value of $S_3$ is 1.4584 which results in $\delta^l = 14.414$ and $\delta^u = 28.828$ using a second set of Bloomberg market data (for 03/20/20) along with the first set for 04/20/20. The full range of Wasserstein radii $\delta$ is given in Table 10. \par
\begin{table}[H]
\begin{center}
\caption{BCVA Wasserstein Radii}
\begin{tabular}{ |c|c|c|c|c|c|c| }
 \hline
Percentage of $\delta^u$ & 50\% & 60\% & 70\% & 80\% & 90\% & 100\% \\
 \hline
W Radius delta & 14.41 & 17.30 & 20.18 & 23.06 & 25.95 & 28.83 \\
 \hline
\end{tabular}
\end{center}
\end{table}
\noindent Matlab plots characterizing the BCVA positive and negative exposure profiles and trajectory of worst case BCVA as a function of Wasserstein radius are shown in Figures 1,2,3. \par


\begin{figure}[!htb]
\caption{Swaps Portfolio Positive Exposure Profiles}
\centerline{\scalebox{0.4}[0.2]{\includegraphics{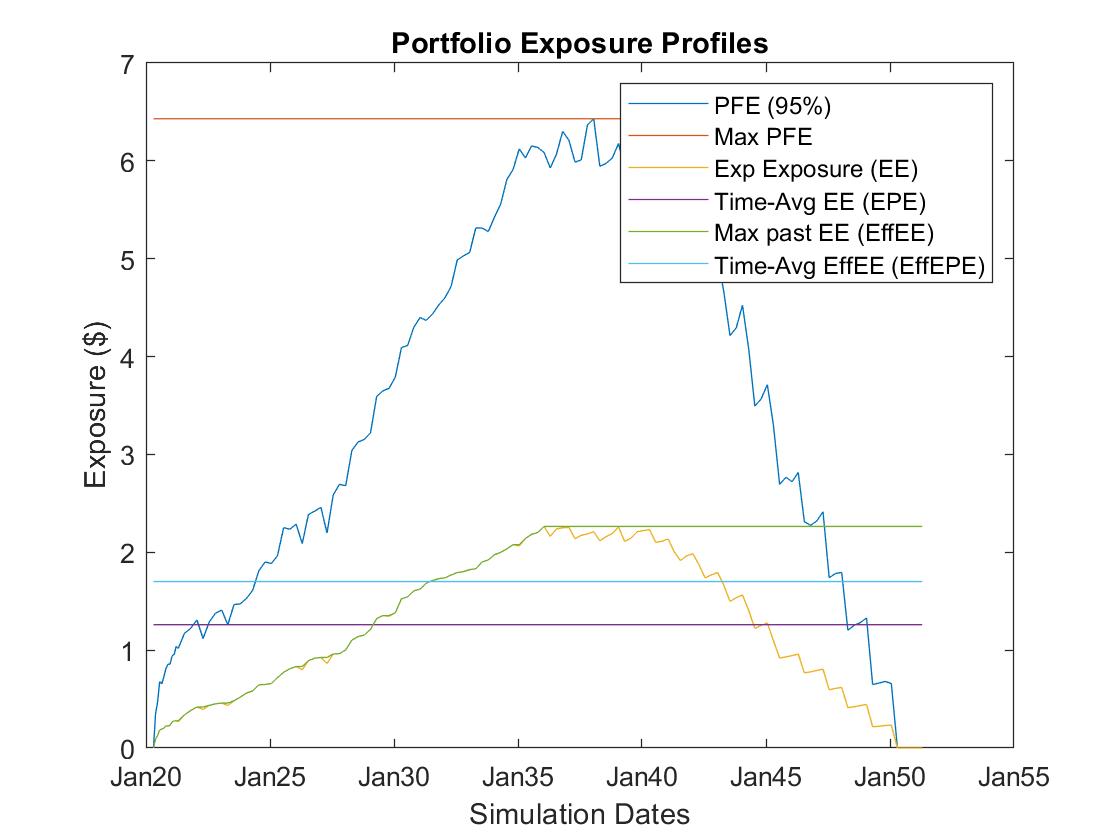}}}
\end{figure}

\begin{figure}[!htb]
\caption{Swaps Portfolio Negative Exposure Profiles}
\centerline{\scalebox{0.4}[0.2]{\includegraphics{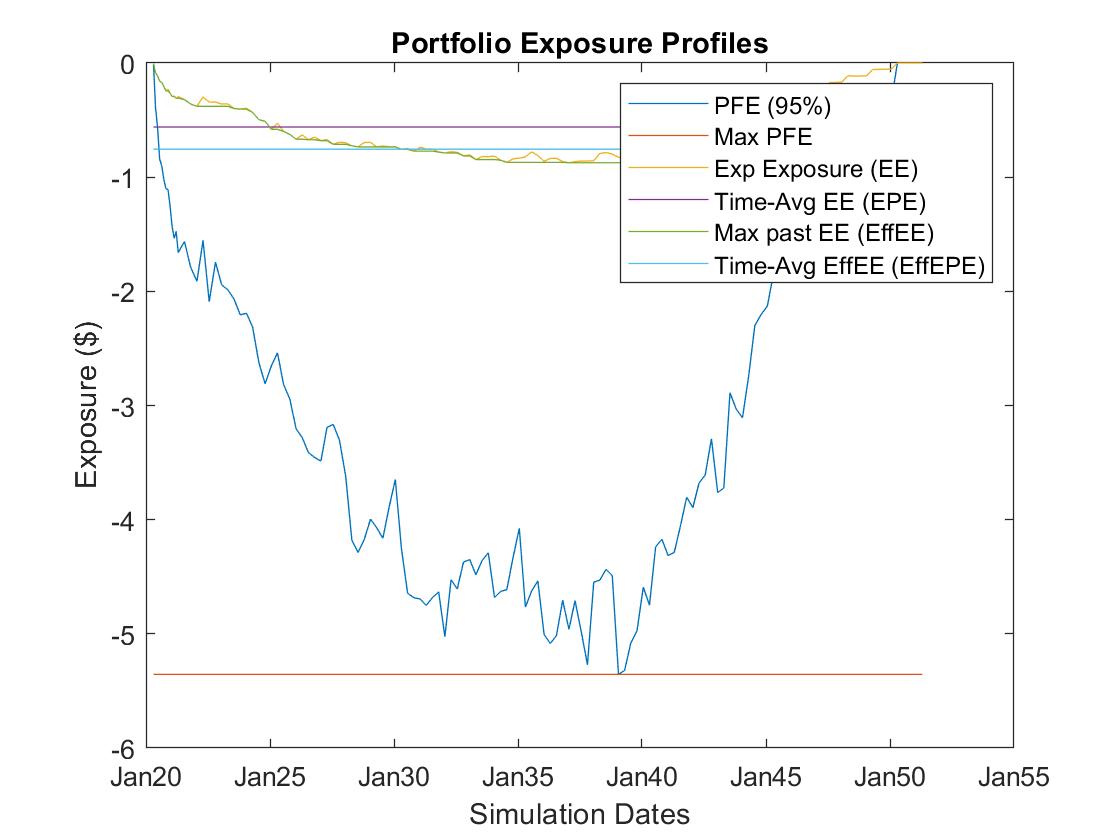}}}
\end{figure}

\begin{figure}[!htb]
\caption{Swaps Portfolio Worst Case BCVA Profile}
\centerline{\scalebox{0.4}[0.2]{\includegraphics{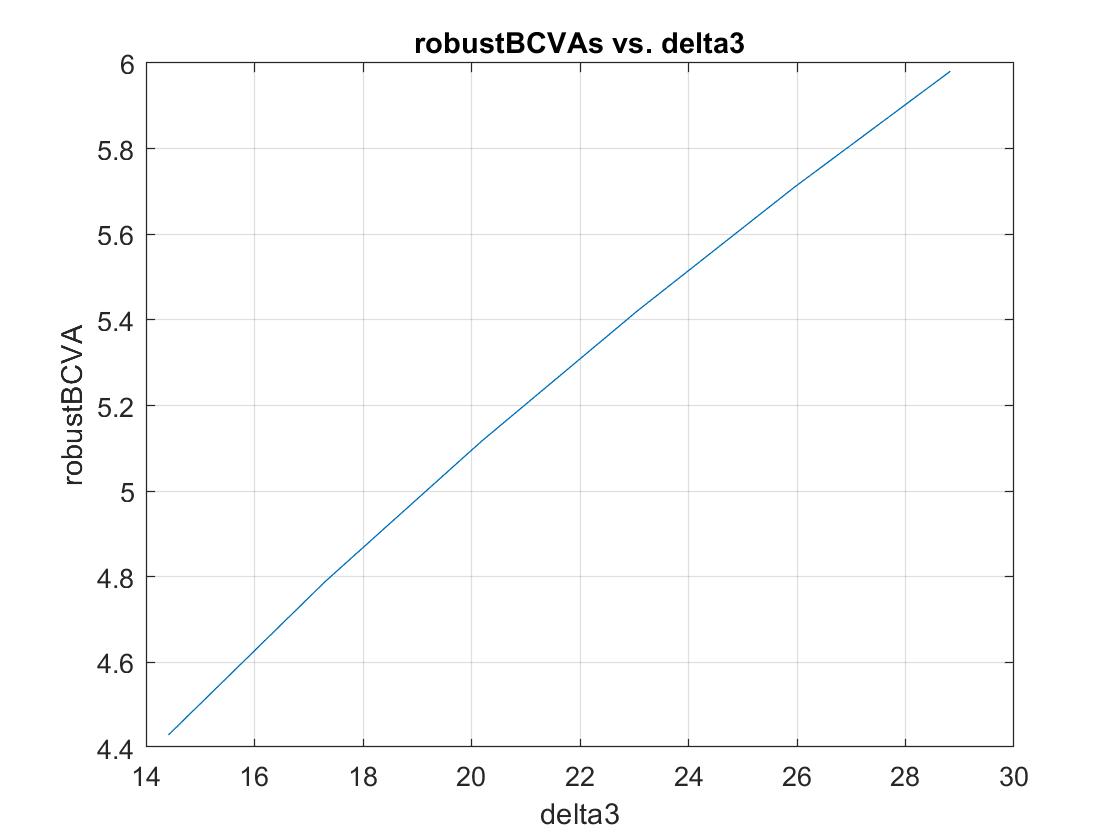}}}
\end{figure}


The baseline BCVA for this portfolio is approximately 160k USD and represents the dot product of the discounted positive portfolio exposure profile times counterparty default probability plus dot product of the discounted negative portfolio exposure times firm default probability. The worst case BCVA curve is shown in Figure 3. The worst case CVA curve ranges from 69\% to 93\% the size of Max PFE (Potential Future Exposure) which is equal to 6.43mm USD (see Figure 1), for Wasserstein radii $\delta$ given in Table 10. So the takeaway here is worst case BCVA can be a significant percentage of PFE for swap portfolios with low risk counterparty default curves (investment grade). \par

The worst case distribution for $\delta^u$ is shown in Figures 4 and 5. The first plot shows the exposures $\{x^*_i\}$ and the second plot shows the joint distribution of counterparty and firm default times $\{y^{c*}_i,y^{f*}_i\}$. Default times beyond the portfolio maturity date denote no default prior to portfolio maturity for those simulation paths. This results in higher contours in the back row.

\begin{figure}[!htb]
\caption{Worst Case Exposures}
\centerline{\scalebox{0.4}[0.2]{\includegraphics{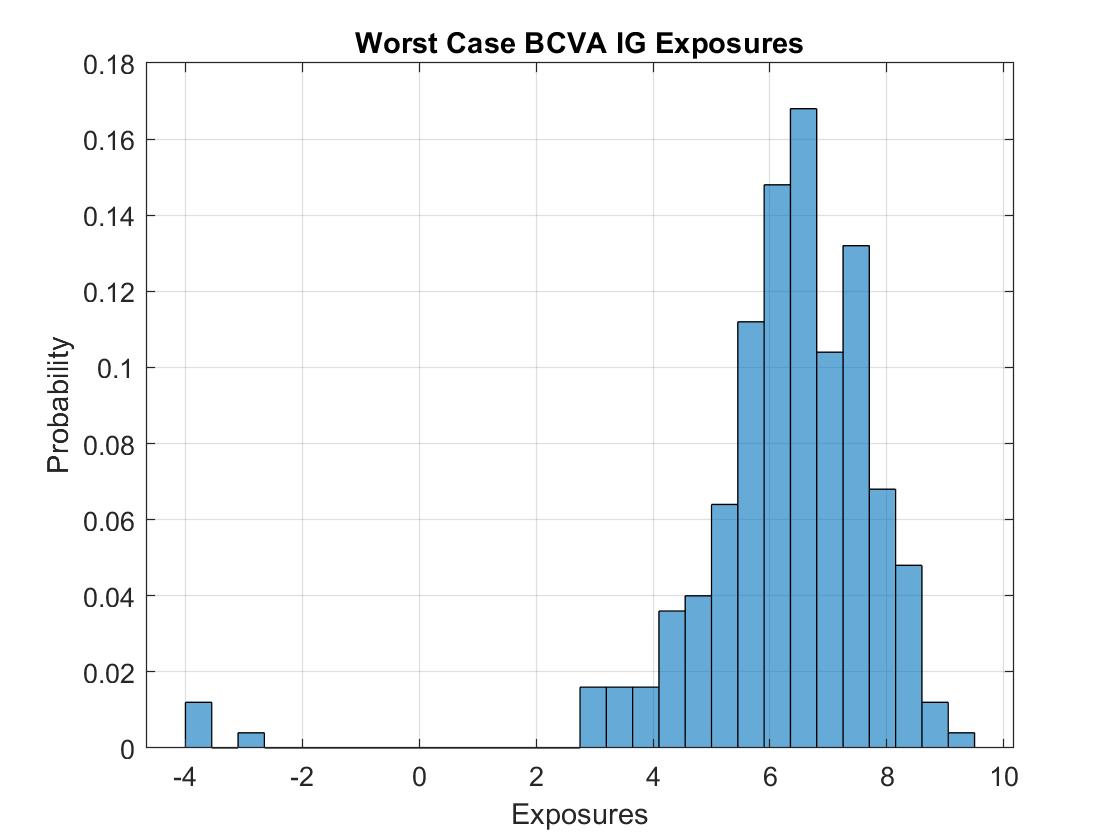}}}
\end{figure}

\begin{figure}[!htb]
\caption{Worst Case Default Times}
\centerline{\scalebox{0.4}[0.2]{\includegraphics{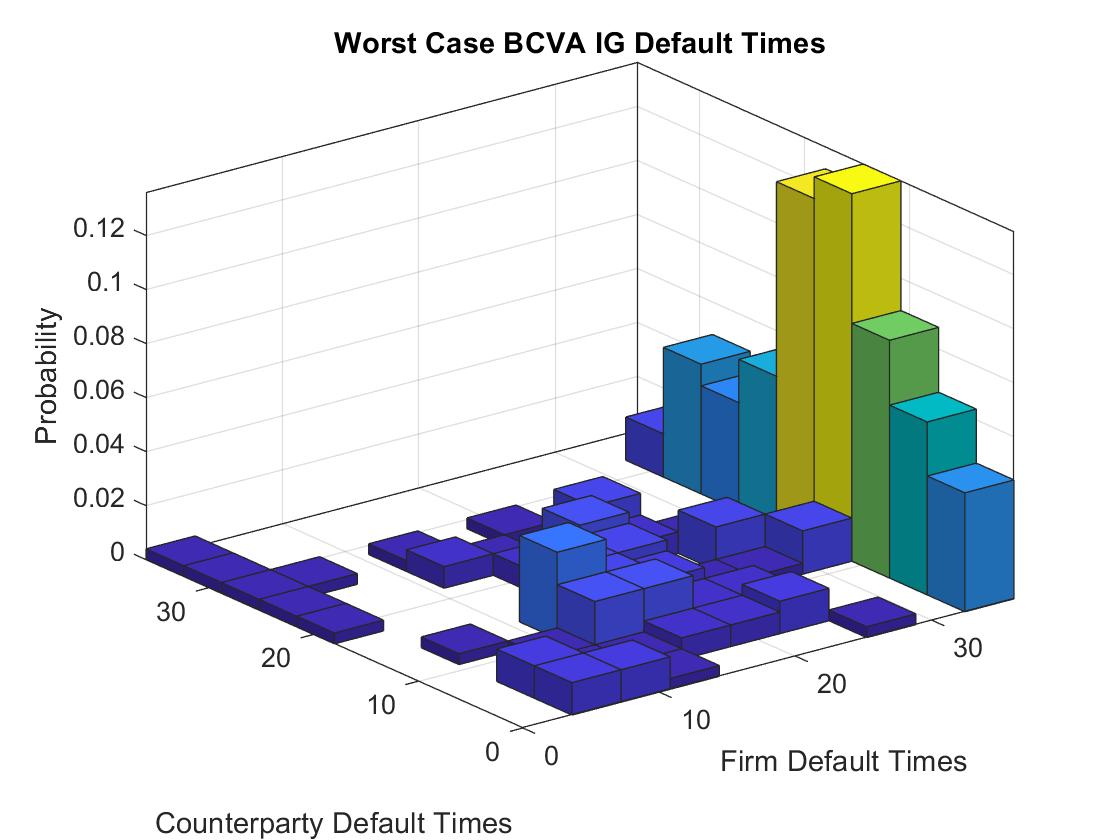}}}
\end{figure}



\subsubsection{High Yield Counterparty and Investment Grade Firm}
The swaps portfolio shown in Table 7 is used for this analysis. The portfolio consists of ten par coupon interest rate swaps, with a mix of receving fixed and paying fixed swaps at different maturities. The high yield counterparty credit spreads are set as in Table 5. The investment grade firm credit spreads are set to a constant 100 basis points. The calibrated value of $S_3$ is 1.4584 which results in $\delta^l = 14.45$ and $\delta^u = 28.90$ using a second set of Bloomberg market data (for 03/20/20) along with the first set for 04/20/20. The full range of Wasserstein radii $\delta$ is given in Table 11. \par
\begin{table}[H]
\begin{center}
\caption{BCVA Wasserstein Radii}
\begin{tabular}{ |c|c|c|c|c|c|c| }
 \hline
Percentage of $\delta^u$ & 50\% & 60\% & 70\% & 80\% & 90\% & 100\% \\
 \hline
W Radius delta & 14.45 & 17.34 & 20.23 & 23.12 & 26.01 & 28.90 \\
 \hline
\end{tabular}
\end{center}
\end{table}
\noindent Matlab plots characterizing the BCVA positive and negative exposure profiles and trajectory of worst case BCVA as a function of Wasserstein radius are shown in Figures 6,7,8. \par


\begin{figure}[!htb]
\caption{Swaps Portfolio Positive Exposure Profiles}
\centerline{\scalebox{0.4}[0.2]{\includegraphics{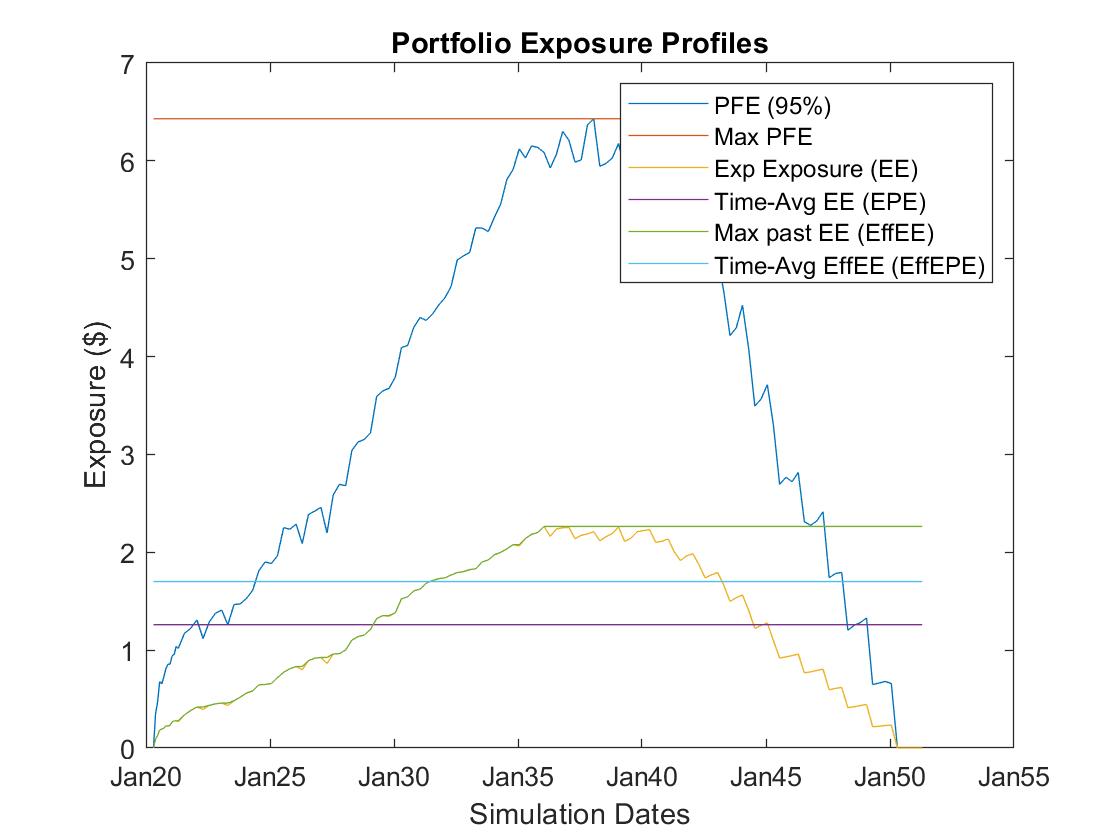}}}
\end{figure}

\begin{figure}[!htb]
\caption{Swaps Portfolio Negative Exposure Profiles}
\centerline{\scalebox{0.4}[0.2]{\includegraphics{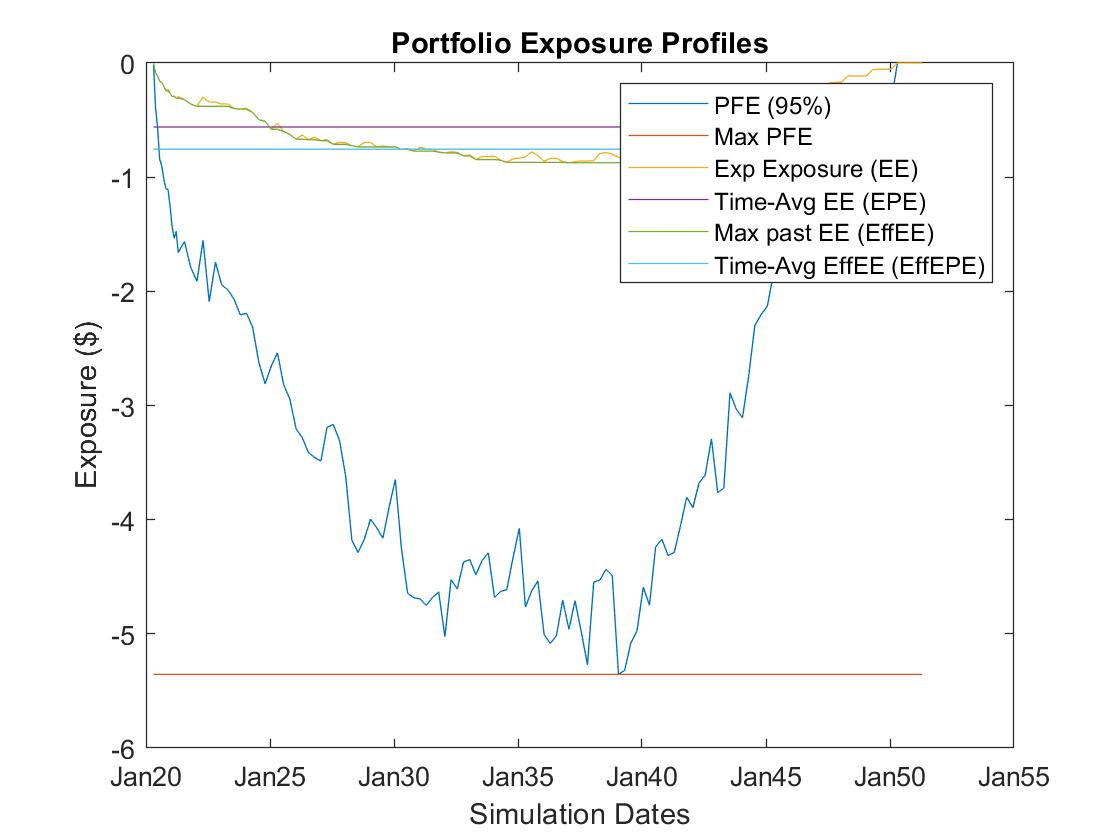}}}
\end{figure}

\begin{figure}[!htb]
\caption{Swaps Portfolio Worst Case BCVA Profile}
\centerline{\scalebox{0.4}[0.2]{\includegraphics{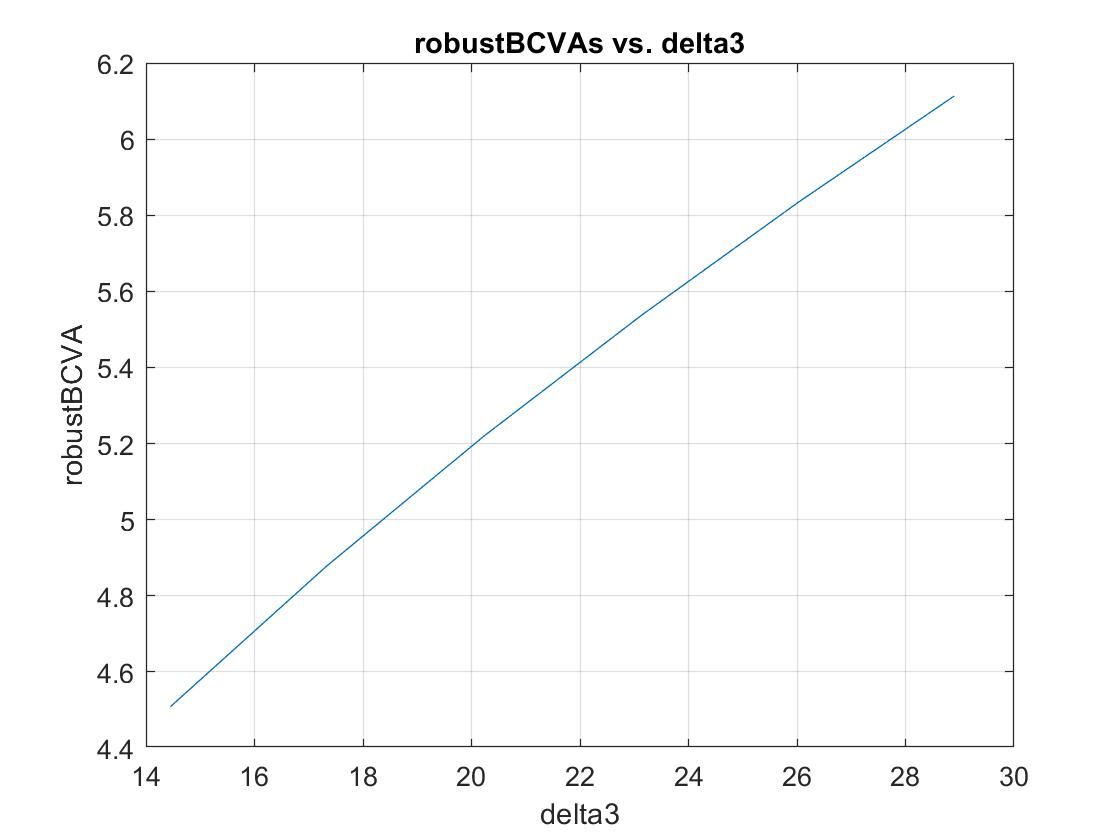}}}
\end{figure}

The baseline BCVA for this portfolio is approximately 106k USD and represents the dot product of the discounted positive portfolio exposure profile times counterparty default probability plus dot product of the discounted negative portfolio exposure times firm default probability. The worst case BCVA curve is shown in Figure 8. Note that for this problem instance, the worst case BCVA results for high yield counterparty credit are similar to the previous subsection, for investment grade counterparty credit. Note the worst case BCVA ranges from 70\% to 95\% the size of Max PFE (Potential Future Exposure), which is equal to 6.43mm USD (see Figure 6), for Wasserstein radii $\delta$ given in Table 11. So the takeaway here is worst case BCVA can be a significant percentage of PFE for swap portfolios with high yield counterparty default curves as well. \par

\begin{figure}[!htb]
\caption{Worst Case Exposures}
\centerline{\scalebox{0.4}[0.2]{\includegraphics{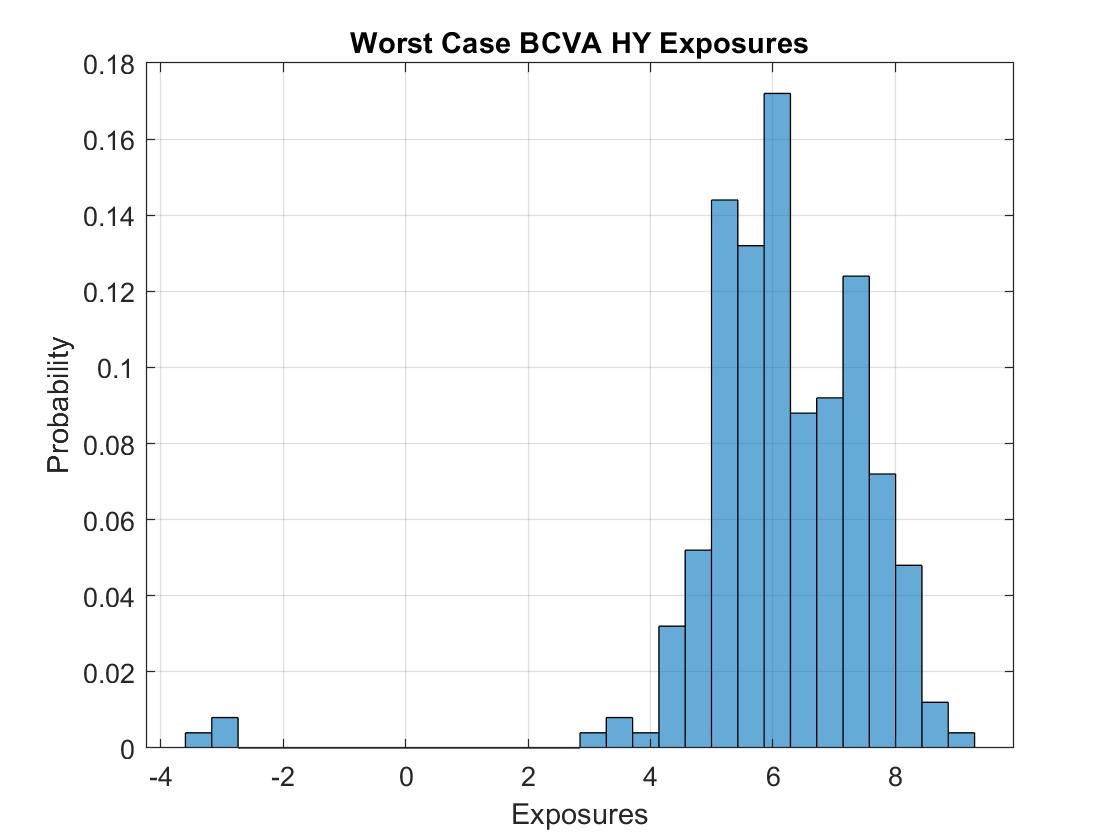}}}
\end{figure}

\begin{figure}[!htb]
\caption{Worst Case Default Times}
\centerline{\scalebox{0.4}[0.2]{\includegraphics{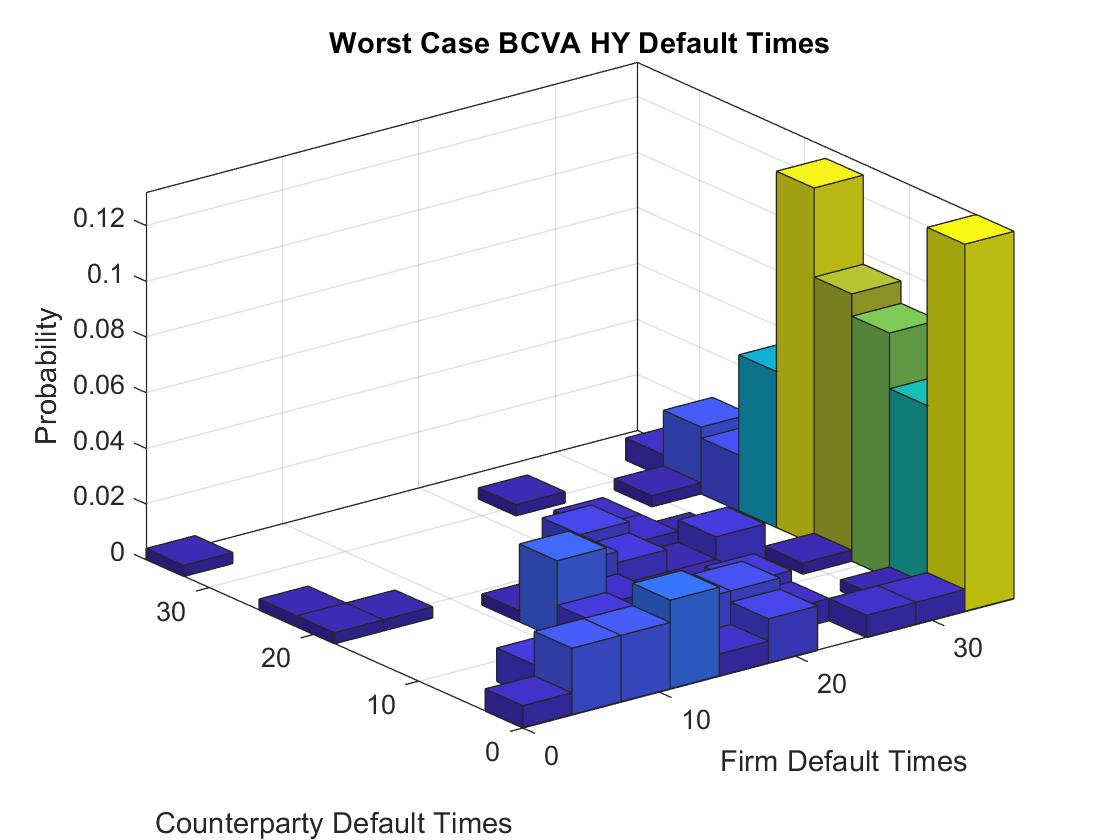}}}
\end{figure}

The worst case distribution for delta value $\delta^u$ is shown in Figures 9 and 10. The first plot shows the exposures $\{x^*_i\}$ and the second plot shows the joint distribution of counterparty and firm default times $\{y^{c*}_i,y^{f*}_i\}$. Default times beyond the portfolio maturity date denote no default prior to portfolio maturity for those simulation paths. This results in higher contours in the joint density plot in the back row. Higher counterparty credit spreads lead to earlier counterparty default times.


\subsection{FVA}
\subsubsection{Investment Grade Counterparty and Firm}

The swaps portfolio shown in Table 8 is used for this analysis. The portfolio consists of ten interest rate swaps, with a mix of receving fixed and paying fixed swaps at different maturities. Capping maturities at 10y introduces some positive NPV to this portfolio. The investment grade firm and counterparty funding spreads are set as shown in Table 6. The calibrated value of $S_3$ is 0.082 which results in $\delta^l = 0.387$ and $\delta^u = 0.774$ using a second set of Bloomberg market data (for 03/20/20) along with the first set for 04/20/20. The full range of Wasserstein radii $\delta$ is given in Table 12. \par
\begin{table}[h]
\begin{center}
\caption{FVA Wasserstein Radii}
\begin{tabular}{ |c|c|c|c|c|c|c| }
 \hline
Percentage of $\delta^u$ & 0.50 & 0.60 & 0.70 & 0.80 & 0.90 & 1.0 \\
 \hline
W Radius delta & 0.387 & 0.464 & 0.542 & 0.560 & 0.697 & 0.774 \\
 \hline
\end{tabular}
\end{center}
\end{table}

\noindent Matlab plots characterizing the FVA positive and negative exposure profiles and trajectory of worst case FVA as a function of Wasserstein radius are shown  in Figures 13,14,15. \par


\begin{figure}[!htb]
\caption{Swaps Portfolio Positive Exposure Profiles}
\centerline{\scalebox{0.4}[0.2]{\includegraphics{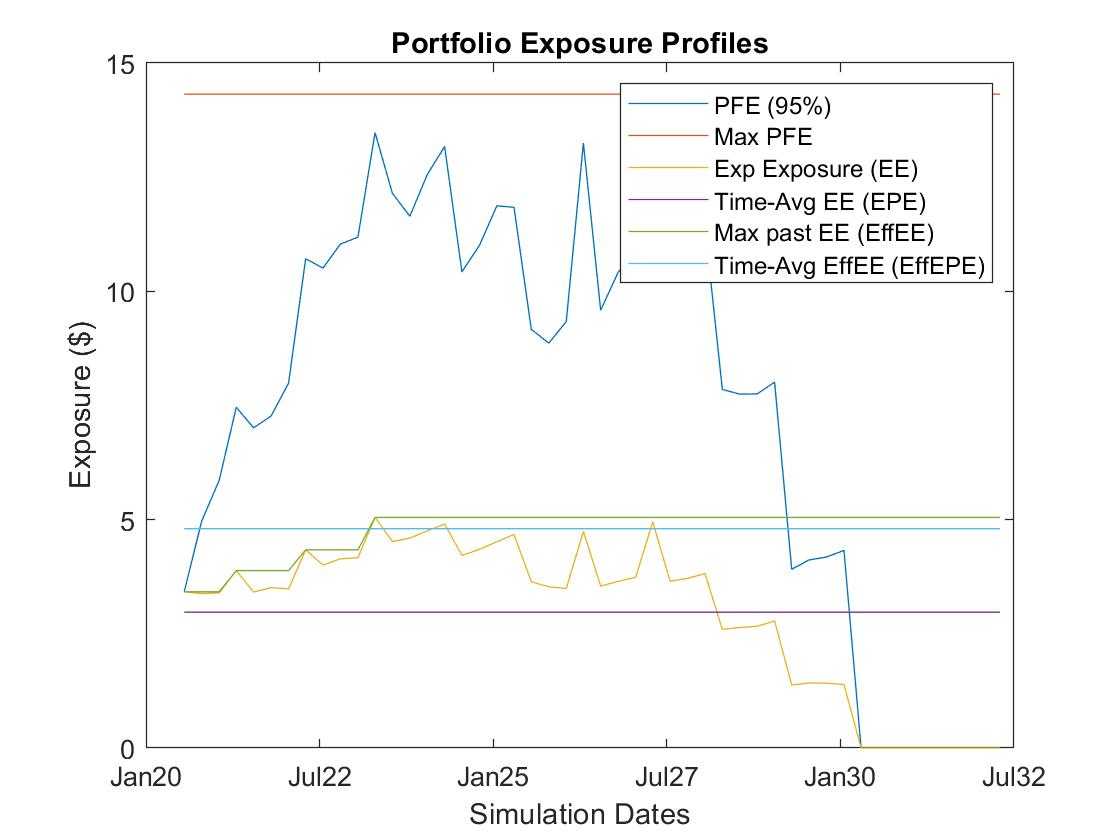}}}
\end{figure}

\begin{figure}[!htb]
\caption{Swaps Portfolio Negative Exposure Profiles}
\centerline{\scalebox{0.4}[0.2]{\includegraphics{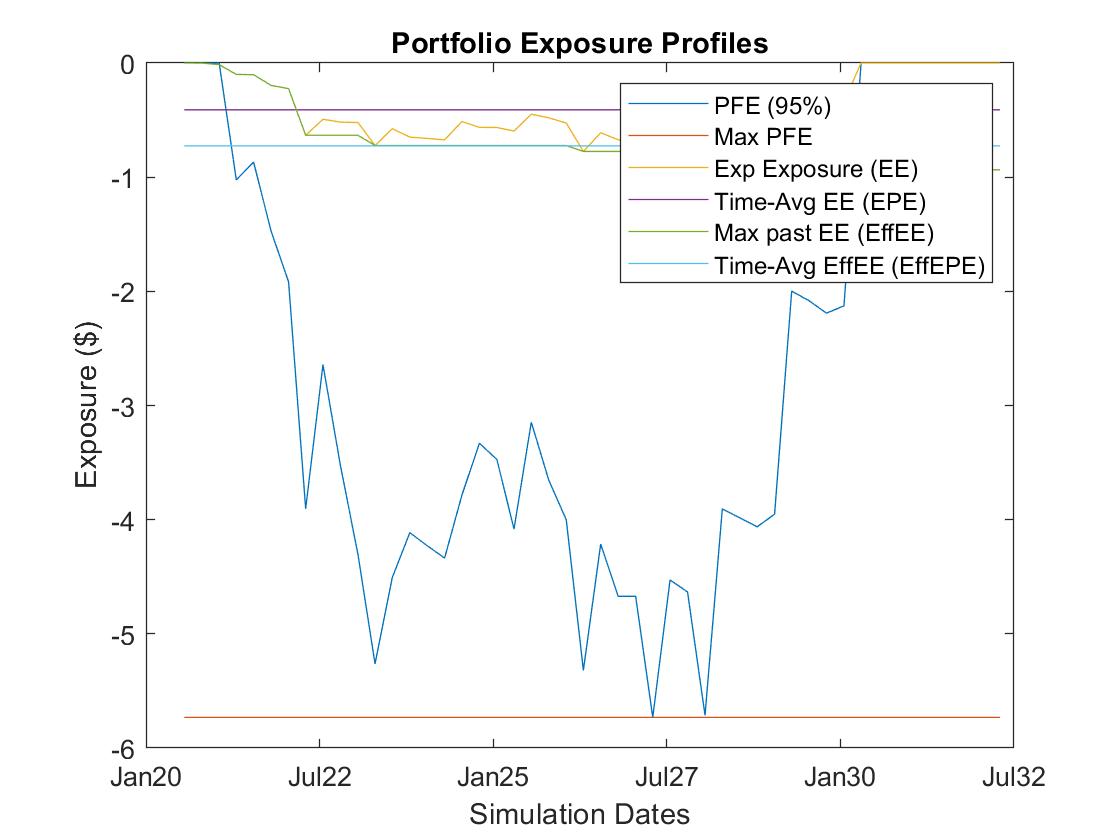}}}
\end{figure}

\begin{figure}[!htb]
\caption{Swaps Portfolio IG FCA Exposure Profiles}
\centerline{\scalebox{0.4}[0.2]{\includegraphics{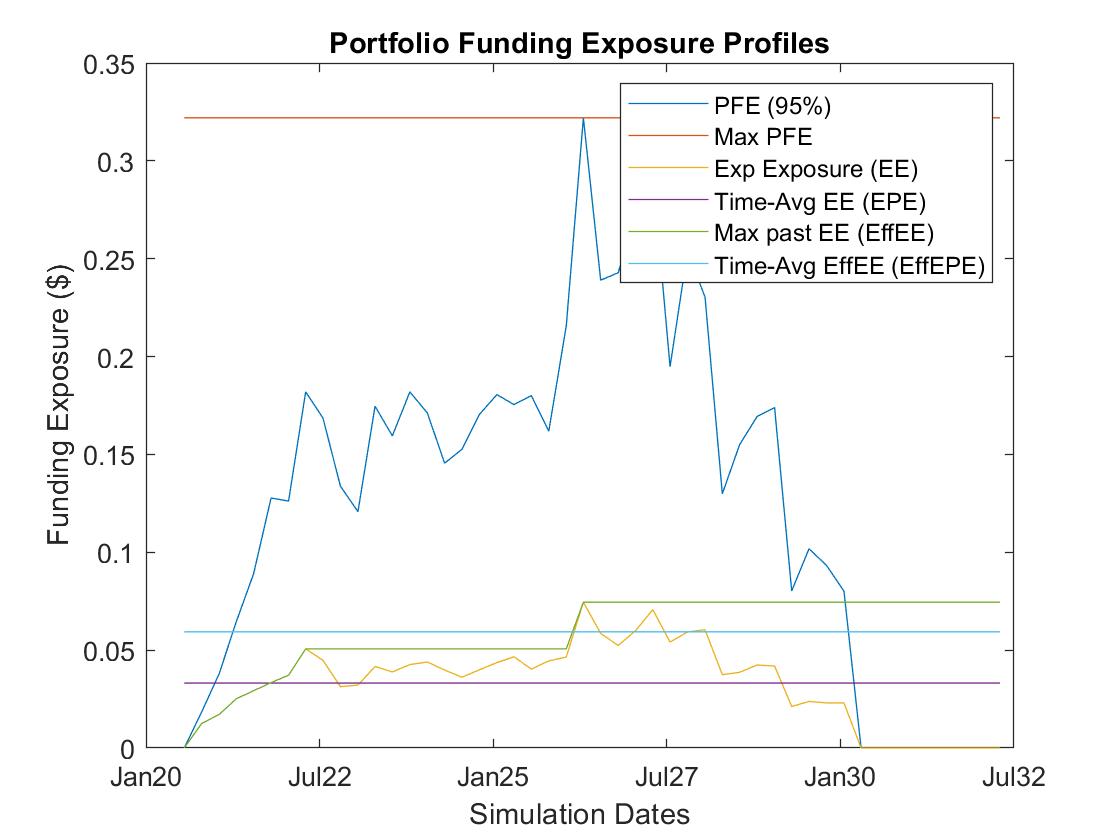}}}
\end{figure}

\begin{figure}[!htb]
\caption{Swaps Portfolio IG FBA Exposure Profiles}
\centerline{\scalebox{0.4}[0.2]{\includegraphics{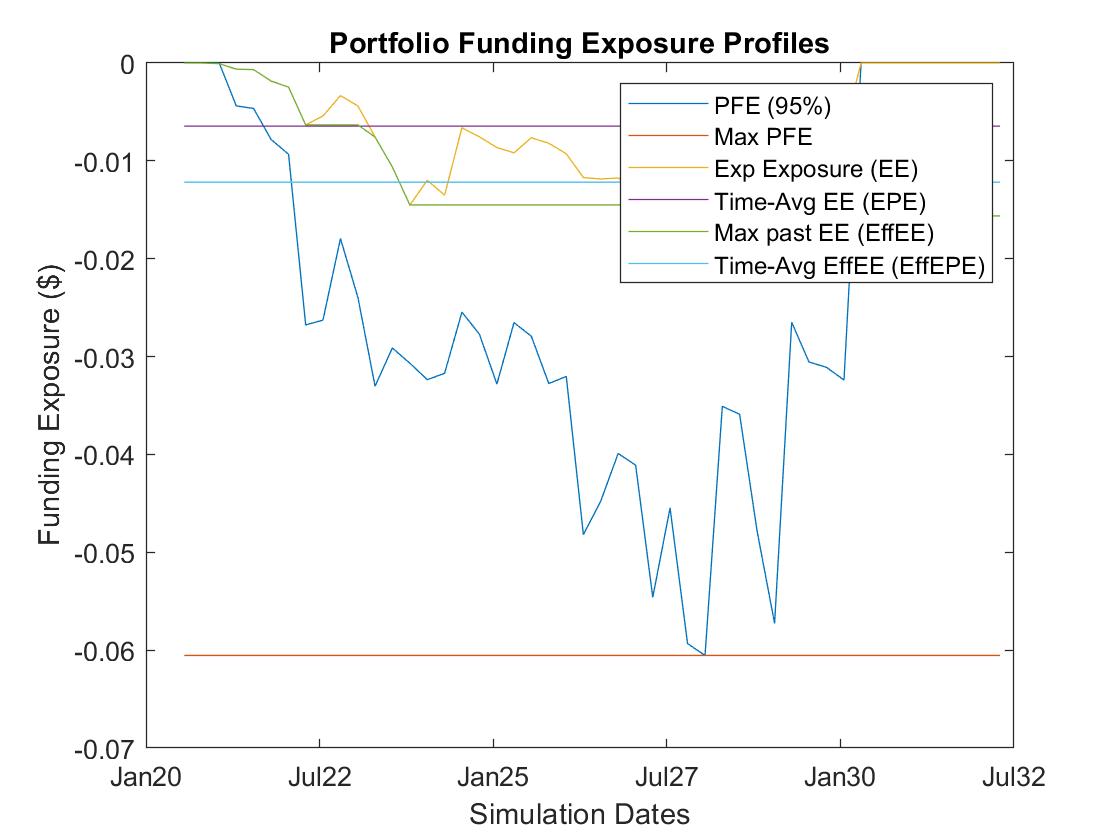}}}
\end{figure}

\begin{figure}[!htb]
\caption{Swaps Portfolio Worst Case IG FVA Profile}
\centerline{\scalebox{0.4}[0.2]{\includegraphics{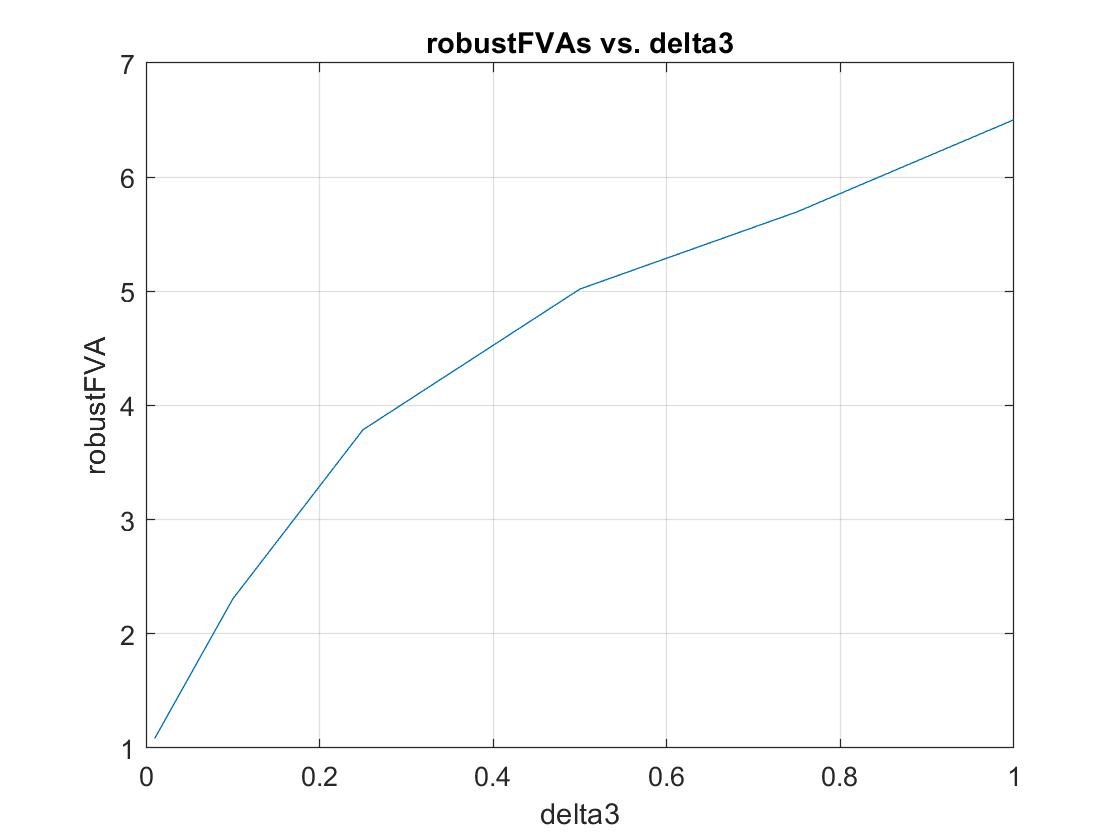}}}
\end{figure}

The baseline FVA for this portfolio is 240k USD and represents the dot product of the discounted portfolio FCA exposure profile times joint survival probability plus dot product of the discounted portfolio FBA exposure times joint survival probability. The worst case FVA curve is shown in Figure 15. For Wasserstein radius $\delta^l = 0.387$, the worst case FVA is approximately 4.55, or 3.0 times the size of integrated FCA PFE which is about 1.54. For Wasserstein radius $\delta^u = 0.774$, the worst case FVA is approximately 5.72, or 3.7 times the size of integrated FCA PFE. In this problem instance, worst case FVA is a multiple of integrated FCA PFE exposure, so quite significant. \par

\begin{figure}[!htb]
\caption{Worst Case Exposures}
\centerline{\scalebox{0.4}[0.2]{\includegraphics{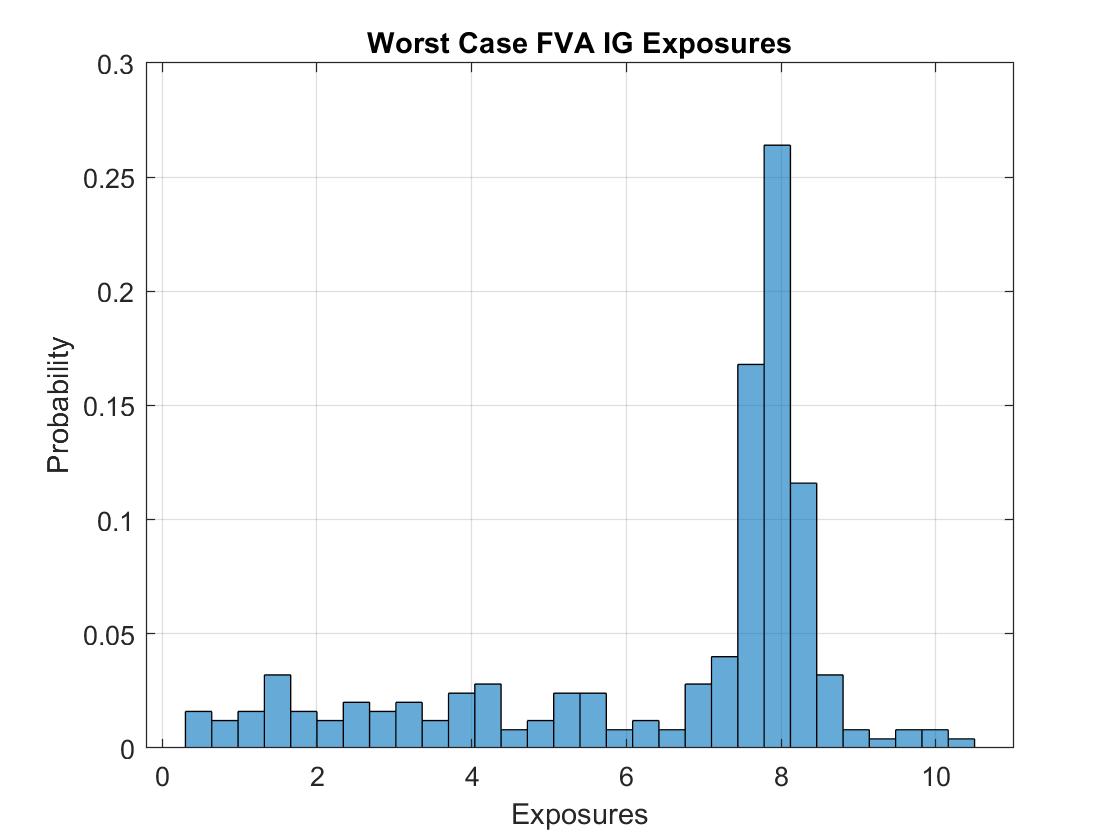}}}
\end{figure}

\begin{figure}[!htb]
\caption{Worst Case Survival Times}
\centerline{\scalebox{0.4}[0.2]{\includegraphics{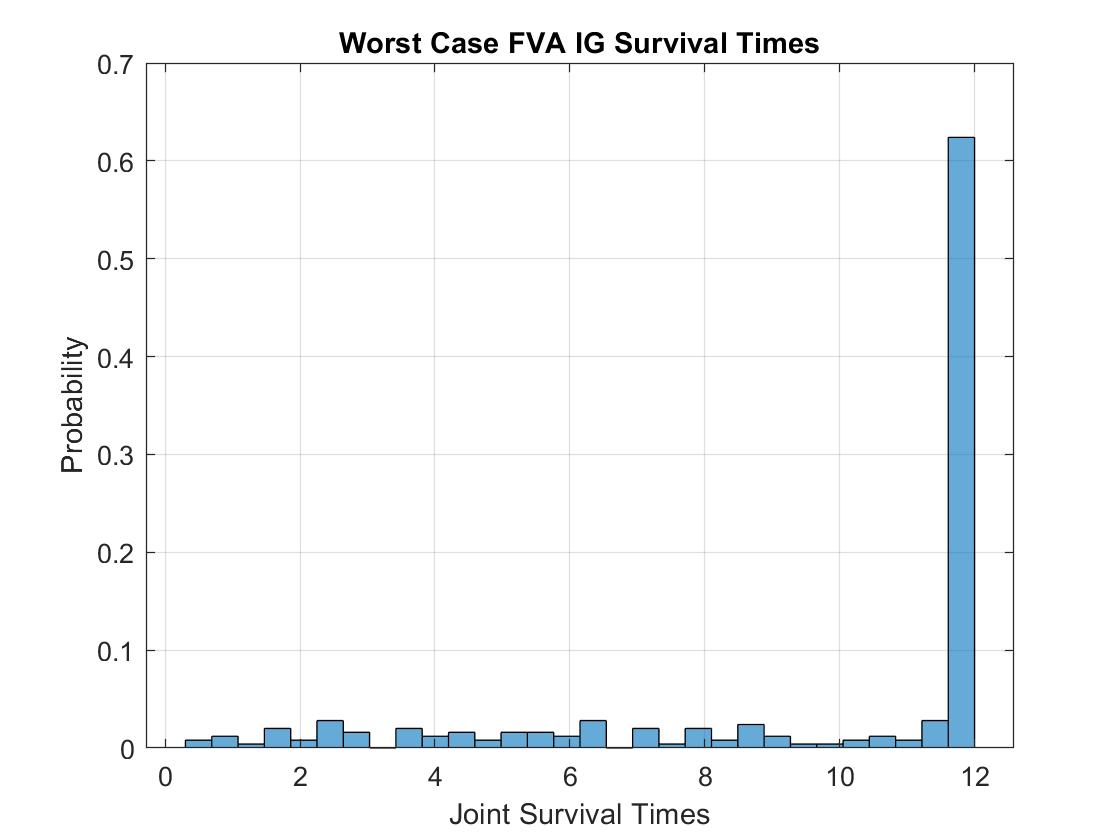}}}
\end{figure}

The worst case distribution for delta value $\delta^u$ is shown in Figures 16 and 17. The first plot shows the exposures $\{z^*_i\}$ and the second plot shows the joint distribution of counterparty and firm survival times $\{y^{cf*}_i\}$. Survival times beyond the portfolio maturity date denote no defaults prior to portfolio maturity for those simulation paths. 


\subsubsection{High Yield Counterparty and Investment Grade Firm}
The swaps portfolio shown in Table 8 is used for this analysis. The portfolio consists of ten par coupon interest rate swaps, with a mix of receving fixed and paying fixed swaps at different maturities. The high yield firm and counterparty funding spreads are set as shown in Table 6. The high yield counterparty credit spreads are set as shown in Table 5. The investment grade firm credit spreads are set to a constant 100 basis points. The calibrated value of $S_3$ is 0.2898 which results in $\delta^l = 1.935$ and $\delta^u = 3.87$ using a second set of Bloomberg market data (for 03/20/20) along with the first set for 04/20/20. The full range of Wasserstein radii $\delta$ is given in Table 13. Matlab plots characterizing the FVA positive and negative exposure profiles and trajectory of worst case FVA as a function of Wasserstein radius are shown  in Figures 20,21,22. \par

\begin{table}[h]
\begin{center}
\caption{FVA Wasserstein Radii}
\begin{tabular}{ |c|c|c|c|c|c|c| }
 \hline
Percentage of $\delta^u$ & 0.50 & 0.60 & 0.70 & 0.80 & 0.90 & 1.0 \\
 \hline
W Radius delta & 1.935 & 2.322 & 2.709 & 3.096 & 3.483 & 3.87 \\
 \hline
\end{tabular}
\end{center}
\end{table}


\begin{figure}[!htb]
\caption{Swaps Portfolio Positive Exposure Profiles}
\centerline{\scalebox{0.4}[0.2]{\includegraphics{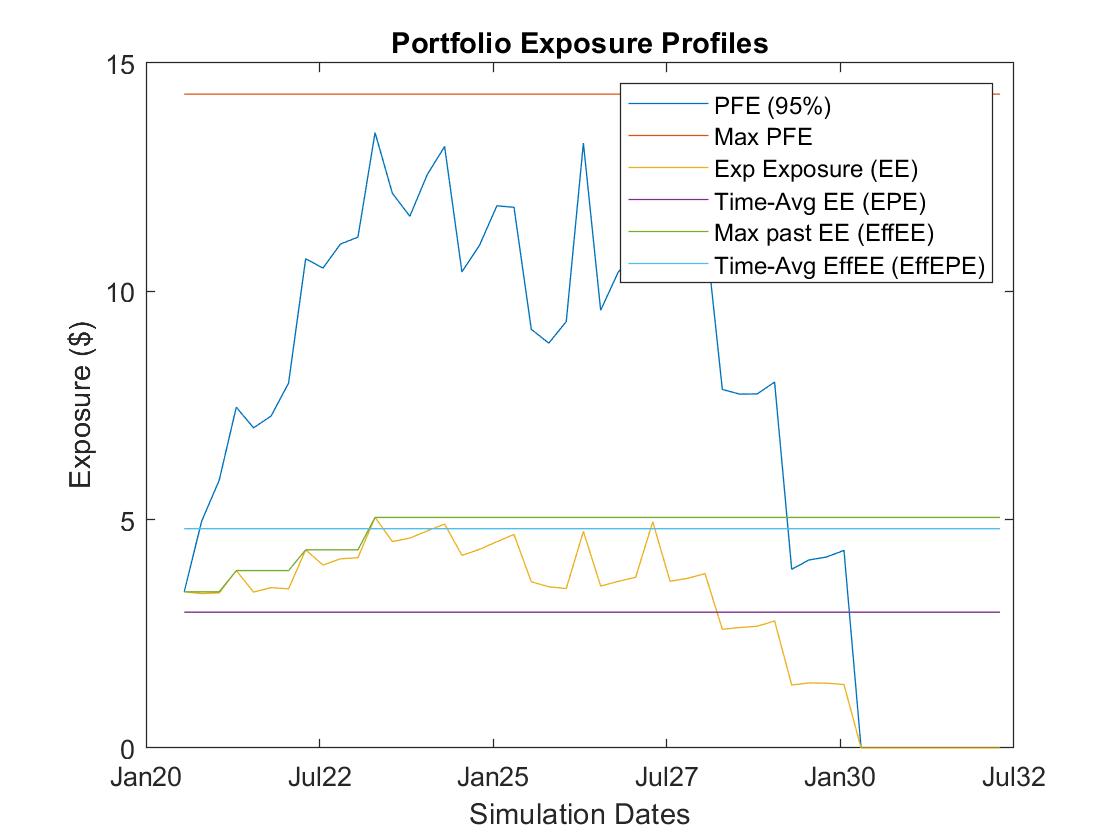}}}
\end{figure}

\begin{figure}[!htb]
\caption{Swaps Portfolio Negative Exposure Profiles}
\centerline{\scalebox{0.4}[0.2]{\includegraphics{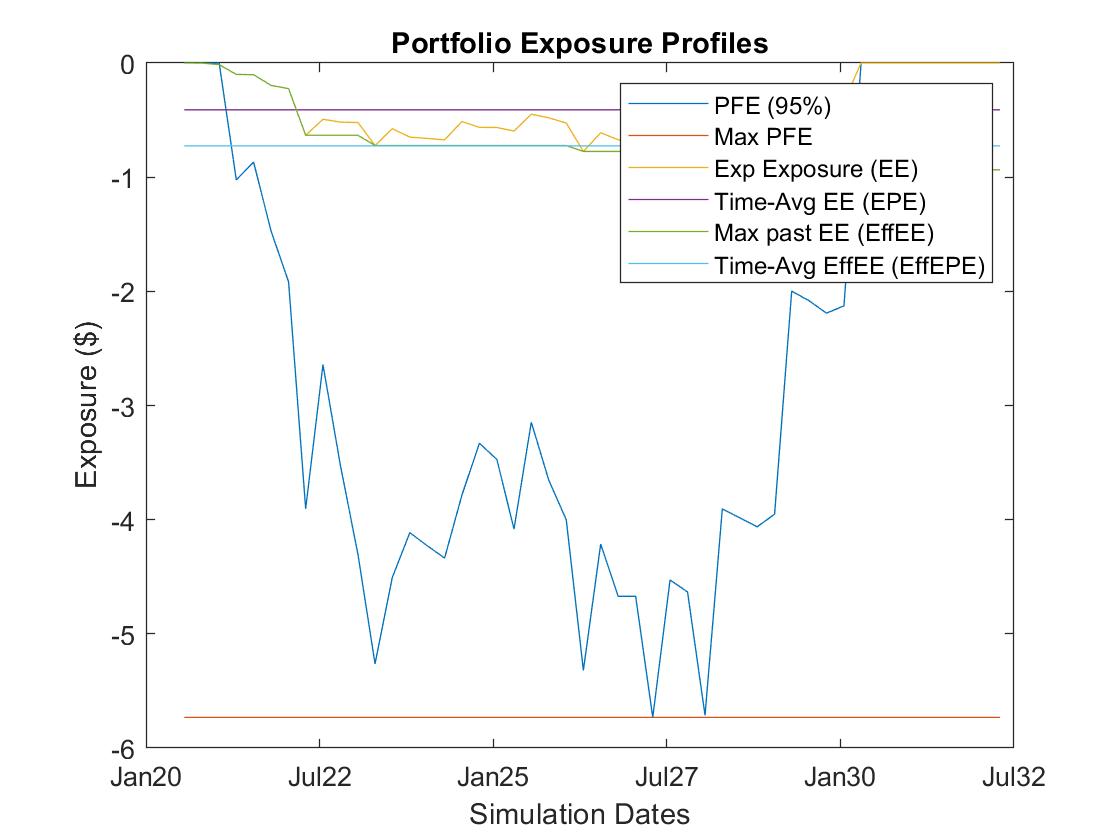}}}
\end{figure}

\begin{figure}[!htb]
\caption{Swaps Portfolio HY FCA Exposure Profiles}
\centerline{\scalebox{0.4}[0.2]{\includegraphics{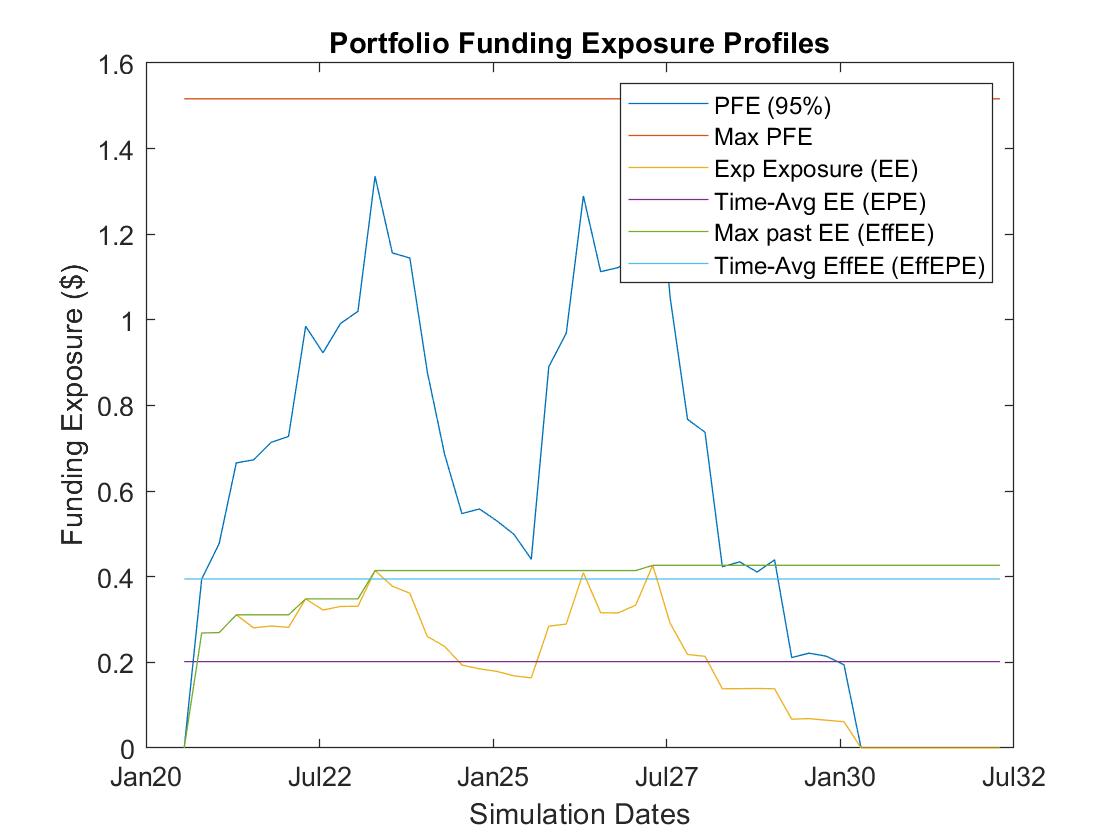}}}
\end{figure}

\begin{figure}[!htb]
\caption{Swaps Portfolio HY FBA Exposure Profiles}
\centerline{\scalebox{0.4}[0.2]{\includegraphics{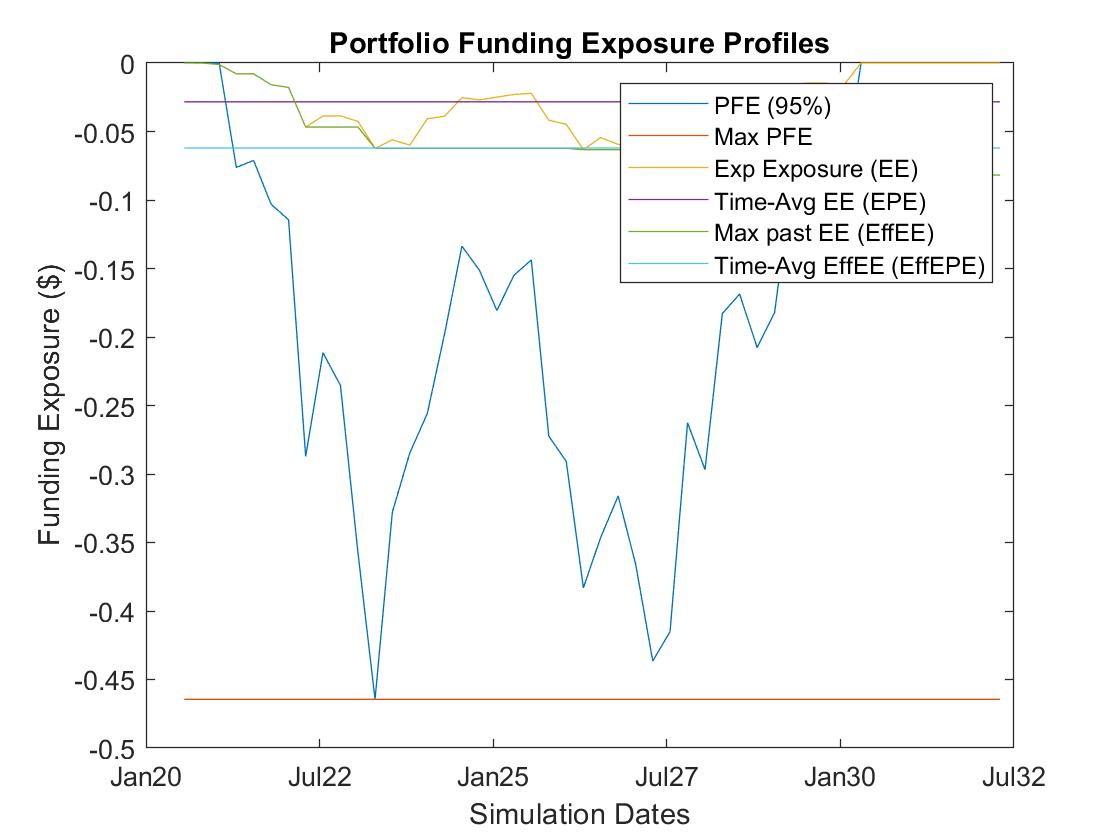}}}
\end{figure}

\begin{figure}[!htb]
\caption{Swaps Portfolio Worst Case HY FVA Profile}
\centerline{\scalebox{0.4}[0.2]{\includegraphics{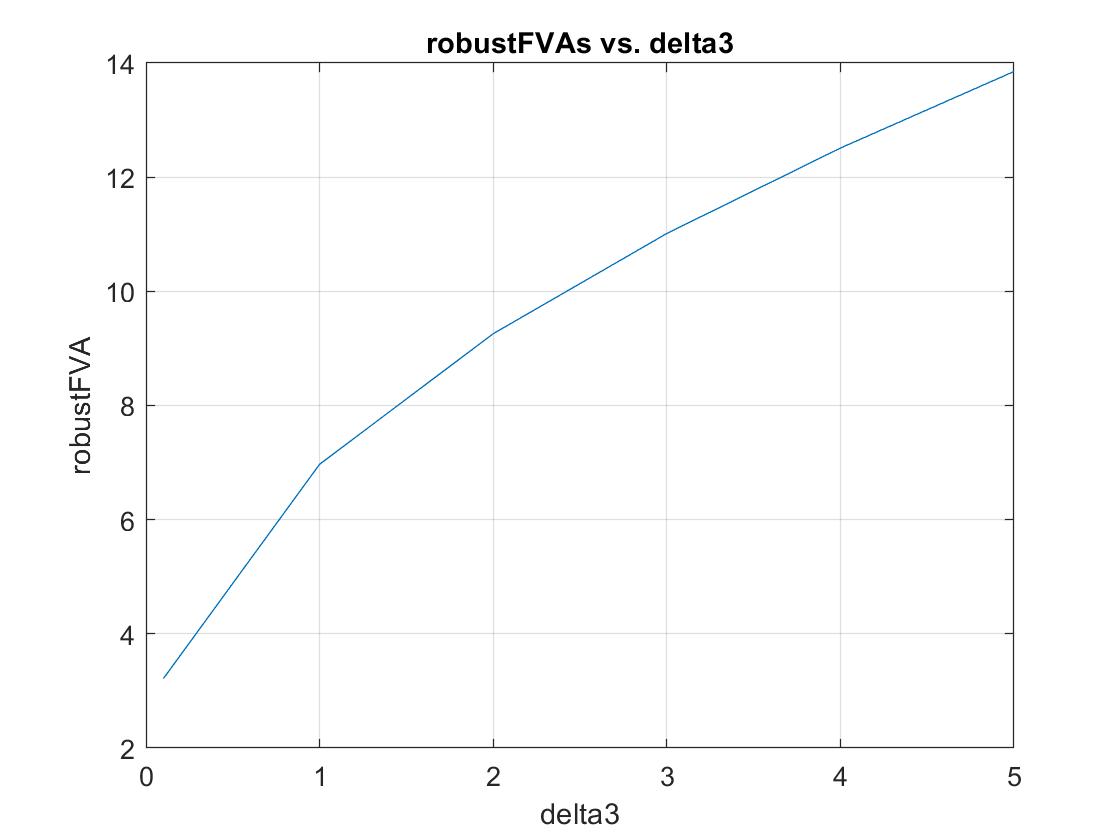}}}
\end{figure}

The baseline FVA for this portfolio is 1.18mm USD and represents the dot product of the discounted portfolio FCA exposure profile times joint survival probability plus dot product of the discounted portfolio FBA exposure times joint survival probability. The worst case FVA curve is shown in Figure 22. For Wasserstein radius $\delta^l = 1.935$, the worst case FVA is approximately 9.35, or 1.31 times the size of integrated FCA PFE which is about 7.127. For Wasserstein radius $\delta^u = 3.87$, the worst case FVA is approximately 12.44, or 1.75 times the size of integrated FCA PFE. In this problem instance, similar to the investment grade example, worst case FVA is a multiple of integrated FCA PFE exposure, so quite significant. \par

\clearpage
\begin{figure}[H]
\caption{Worst Case Exposures}
\centerline{\scalebox{0.4}[0.2]{\includegraphics{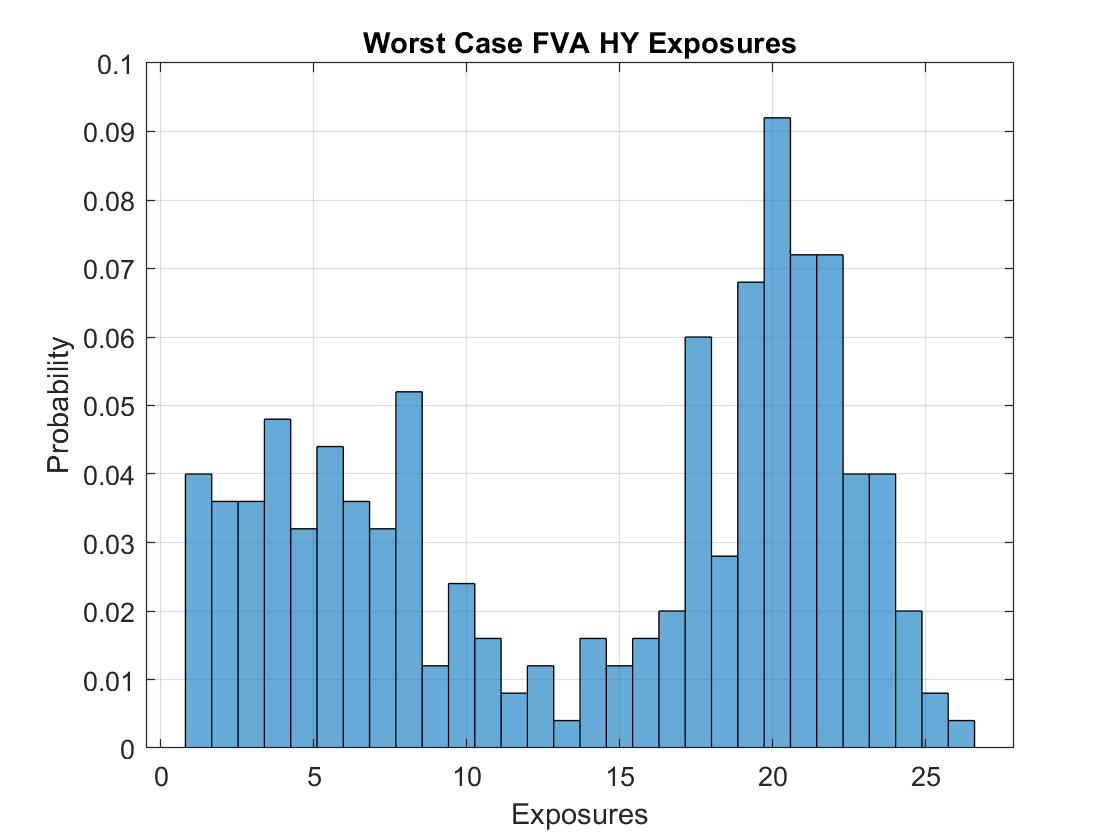}}}
\end{figure}

\begin{figure}[H]
\caption{Worst Case Survival Times}
\centerline{\scalebox{0.4}[0.2]{\includegraphics{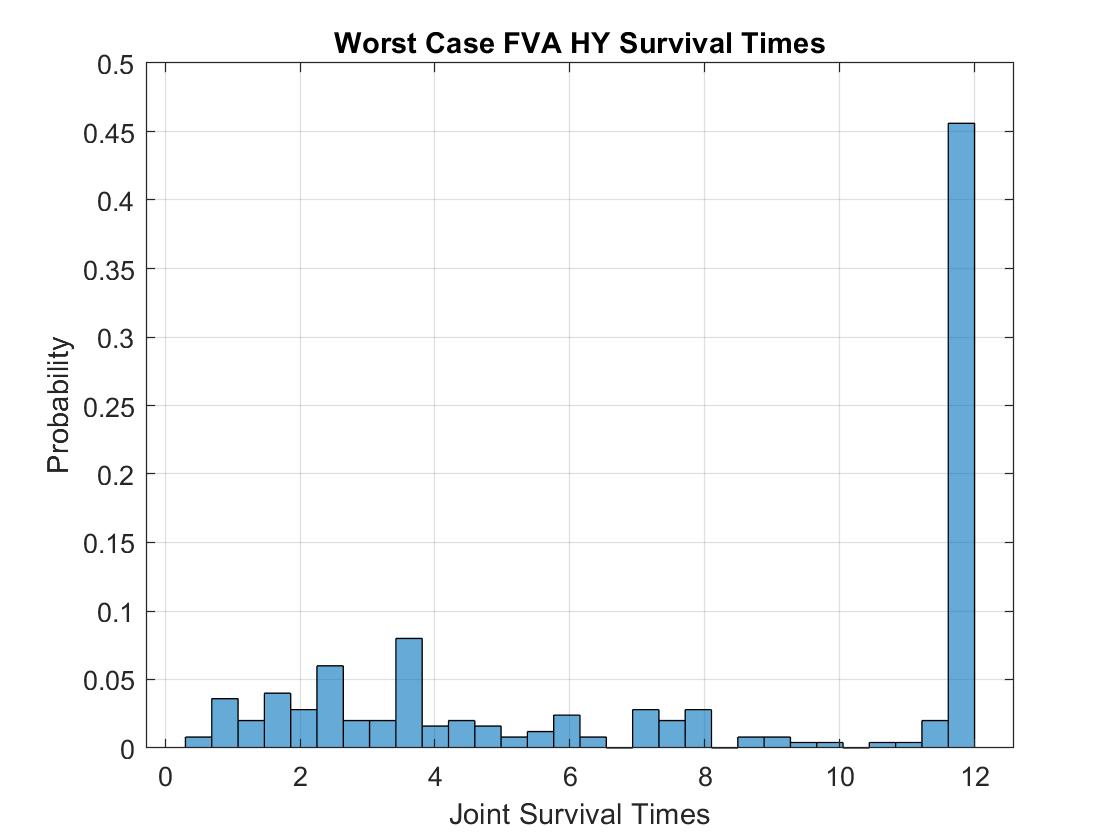}}}
\end{figure}

The worst case distribution for delta value $\delta^u$ is shown in Figures 23 and 24. The first plot shows the exposures $\{z^*_i\}$ and the second plot shows the joint distribution of counterparty and firm survival times $\{y^{cf*}_i\}$. Survival times beyond the portfolio maturity date denote no defaults prior to portfolio maturity for those simulation paths. 

\section{Conclusions and Further Work}
This work has developed theoretical results and investigated calculations of robust CVA, FVA, and wrong way risk for OTC derivatives under distributional uncerainty using Wasserstein distance as an ambiguity measure. The financial market overview, foundational notation, and robust XVA primal problems were introduced in Section 1. Using recent infinite dimensional Lagrangian duality results \citep{blanchetFirst}, the simpler dual formulations and their analytic solutions for BCVA and FVA were derived in Section 2. After that, in Section 3, some computational experiments were conducted to measure the additional XVA charge due to distributional uncertainty for a variety of portfolio and market configurations. Worst case BCVA and FVA were found to be significant relative to their respective PFE profiles in all problem instances. Finally, we conclude with some commentary on directions for further research. \par

One direction for future research, as has been previously discussed, is to extend the problem formulations to include a risk neutral measure constraint in a solvable way. Explicitly adding the constraint would complicate the problem formulations no doubt, so perhaps there is a more tractable indirect approach. Another direction for future research would be to develop (and apply) similar theoretical machinery as used for robust CVA and FVA towards robust KVA (Capital Valuation Adjustment) and MVA (Margin Valuation Adjustment) and wrong way risk in that context. Intuitively, wrong way risk arises in that context when the market cost of capital and/or funding the margin position increases at the same time as the portfolio exposure increases.\par

\section*{Data Availability Statement}
The raw and/or processed data required to reproduce the findings from this research can be obtained from the corresponding author, [D.S.], upon reasonable request.

\section*{Conflict of Interest Statement}
The authors declare they have no conflict of interest.

\section*{Funding Statement}
The authors received no specific funding for this work.

\clearpage
\bibliographystyle{apalike}

\bibliography{RobustXVA}

\clearpage
\end{document}